\documentclass[11pt,double,draftclsnofoot,onecolumn]{IEEEtran}
\usepackage[cmex10]{amsmath}
\usepackage{graphics}
\usepackage{mdwmath,mdwtab,url}
\usepackage[caption=false,font=footnotesize]{subfig}
\usepackage{stfloats}
\usepackage{nomencl}
\usepackage{amssymb}
\usepackage{rotating}
\usepackage{psfrag}
\usepackage{color}
\usepackage[dvipsnames]{xcolor}
\usepackage{hyperref}
\usepackage{cite}
\usepackage{epstopdf}
\newtheorem{theorem}{ {Theorem}}

\newtheorem{proposition}{ {Proposition}}
\newtheorem{definition}{{Definition}}

\newtheorem{remark}{ {Remark}}
\newcommand{\mb}{\mathbb} 
\newcommand{\dv}{\mathbf} 
\newcommand{\mc}{\mathcal} 

\newcommand{\tf}{\textbf}

\newcommand{\tc}{\textcolor}

\renewcommand\footnotemark{}

\begin{document}
\title{Secure Degrees of Freedom of MIMO X-Channels with Output Feedback and Delayed CSIT}

\author{Abdellatif~Zaidi, Zohaib~Hassan~Awan, Shlomo Shamai (Shitz)~\IEEEmembership{Fellow,~IEEE}, Luc~Vandendorpe~\IEEEmembership{Fellow,~IEEE} 

\thanks{Copyright (c) 2013 IEEE. Personal use of this material is permitted. However, permission to use this material for any other purposes must be obtained from the IEEE by sending a request to pubs-permissions@ieee.org.}

\thanks{Abdellatif Zaidi is with Universit\'e Paris-Est Marne-la-Vall\'ee, 77454 Marne-la-Vall\'ee Cedex 2, France. Email: abdellatif.zaidi@univ-mlv.fr}

 \thanks{Zohaib Hassan Awan was with the ICTEAM institute, Universit\'e catholique de Louvain, Louvain-la-Neuve. Email: zohaib.awan@uclouvain.be}

\thanks{Shlomo Shamai (Shitz) is with the Department of Electrical Engineering, Technion Institute of Technology, Technion City, Haifa 32000, Israel. Email: sshlomo@ee.technion.ac.il}

\thanks{Luc Vandendorpe is with the ICTEAM institute (\'{E}cole Polytechnique de Louvain), Universit\'e catholique de Louvain, Louvain-la-Neuve 1348, Belgium. Email: luc.vandendorpe@uclouvain.be}

\thanks{This work has been supported by the European Commission in the framework of the FP7 Network of Excellence in Wireless Communications (NEWCOM\#), and the Concerted Research Action, SCOOP. The authors would also like to thank BELSPO for the support of the IAP BESTCOM  network.}

\thanks{ The results in this work will be presented in part at the IEEE Information Theory Workshop, Sept. 2013.}
}

\maketitle

\begin{abstract}
We investigate the problem of secure transmission over a two-user multi-input multi-output (MIMO) X-channel in which channel state information is provided with one-unit delay to both transmitters (CSIT), and each receiver feeds back its channel output to a different transmitter. We refer to this model as MIMO X-channel with \textit{asymmetric} output feedback and delayed CSIT. The transmitters are equipped with $M$ antennas each, and the receivers are equipped with $N$ antennas each. For this model, accounting for both messages at each receiver, we characterize the optimal sum secure degrees of freedom (SDoF) region. We show that, in presence of asymmetric output feedback and delayed CSIT, the sum SDoF region of the MIMO X-channel is  \textit{same} as the SDoF region of a two-user MIMO BC with $2M$ antennas at the transmitter, $N$ antennas at each receiver and delayed CSIT. This result shows that, upon availability of asymmetric output feedback and delayed CSIT, there is no performance loss in terms of sum SDoF due to the distributed nature of the transmitters. Next, we show that this result also holds if only output feedback is conveyed to the transmitters, but in a \textit{symmetric} manner, i.e., each receiver feeds back its output to both transmitters and no CSIT. We also study the case in which only asymmetric output feedback is provided to the transmitters, i.e., without  CSIT, and derive a lower bound on the sum SDoF for this model. Furthermore, we specialize our results to the case in which there are no security constraints. In particular, similar to the setting with security constraints, we show that the optimal sum DoF region of the $(M,M,N,N)$--MIMO X-channel with asymmetric output feedback and delayed CSIT is same as the DoF region of a two-user MIMO BC with $2M$ antennas at the transmitter, $N$ antennas at each receiver, and delayed CSIT. We illustrate our results with some numerical examples.
\end{abstract}

\section{Introduction}\label{secI}

In modern era, there is a growing requirement for high data rates in wireless networks, in which multiple users communicate with each other over a shared medium. The information transmission by multiple users on a common channel raises an important issue of interference in networks. In existing literature on multi-user channels, such as \cite{J10}, several interference alignment techniques have been proposed. Most of these techniques rely on the availability of perfect channel state information at the transmitting nodes (CSIT). However, because the wireless medium is characterized by its inherent randomness, such an assumption is rather idealistic and is difficult to obtain in practice. In \cite{M-AT12}, Maddah-Ali and Tse study a  multi-input single-output (MISO) broadcast channel with delayed CSI available at the transmitter, from a degrees of freedom (DoF) perspective. They show that delayed (or stale) CSIT is useful, in the sense that it increases the DoF region in comparison with the same MISO setting without any CSIT. The model with delayed CSIT of \cite{M-AT12} has been extended to study a variety of models. These include the two-user MIMO BC \cite{vaze_broadcast}, the three-user MIMO BC \cite{abdoli,vaze_broadcast}, the two-user MIMO interference channel \cite{vaze_int,akbar_int}, and the $K$-user single-input single-output (SISO) interference and X-channels~\cite{javad,TMPS12b}.

In~\cite{JS08}, Jafar and Shamai introduced a two-user X-channel model. The two-user X-channel consists of two transmitters and two receivers, with each transmitter sending two independent messages to both receivers. For this model, the authors establish bounds on the DoF region under the assumption of full CSIT. In~\cite{maleki}, Maleki \textit{et al.} study a two-user SISO X-channel with output feedback provided asymmetrically to the transmitters. They establish a lower bound on the allowed sum DoF. For MIMO X-channels, the setting with no CSIT is studied in~\cite{vaze_no}; the setting with delayed CSIT is studied~\cite{akbar}; and the setting with delayed CSIT and asymmetric noiseless output feedback is studied in~\cite{TMPS12a}, all from a DoF viewpoint. In all these works, a symmetric antenna topology is assumed, with each transmitter being equipped with $M$ antennas and each receiver equipped with $N$ antennas. In \cite{akbar}, it is assumed that each receiver knows the CSI of its own channel and also the past CSI of the channel to the other receiver. Also, the past CSI available at each receiver is provided to the corresponding transmitter over a noiseless link. For this model, the authors establish a lower bound on the sum DoF over all messages in the network (in the rest of this paper, we will refer to this as being the \textit{total} DoF). In \cite{TMPS12a}, Tandon \textit{et al.} study a model which is similar to the one that is investigated in \cite{akbar}, but with additional asymmetric noiseless output feedback from the receivers to the transmitters. In particular, they show that the total DoF of this two-user MIMO X-channel with asymmetric output feedback and delayed CSIT is same as the total DoF of a two-user broadcast channel with delayed CSIT, with $2M$ transmit antennas and $N$ antennas at each receiver. For this model, the availability of the output feedback together with the delayed CSIT help each transmitter reconstruct the information transmitted by the other transmitter. The reader may refer to~\cite{YZ12,MCJ12,SGJ13} for some other related works. 

In his seminal work \cite{wyner}, Wyner introduced a basic information-theoretic model to study security by exploiting the physical layer attributes of the channel. The model consists of a sender which transmits information to a legitimate receiver; and this information is meant to be kept secret from an external wiretapper that overhears the transmission. Wyner's basic setup has been extended to study the secrecy capacity of various multiuser channels, such as the broadcast channel \cite{csiszar,LLPS13}, the multi-antennas wiretap channel~\cite{khisti,oggier,Tie,BLPS09}, the multiple access wiretap channel \cite{tekin,liangpoor,tekin2,Mac_2012,MAC_ieee}, the relay channel \cite{lai,Z_allerton_2010,Z_relay_ieee}, the interference channel \cite{onur_IFC,LYT08} and X networks~\cite{GJ08} (the reader may also refer to \cite{liangbook} for a review of many other related contributions). In~\cite{onur_dof}, the authors study a $K$-user interference channel with security constraints, from a SDoF perspective. Similar to the setting with no security constraints, the SDoF captures the way the spatial multiplexing gain, or secrecy capacity prelog or degrees of freedom, scales asymptotically with the logarithm of the signal-to-noise ratio (SNR). In \cite{ghadamali}, the authors study a $K$-user Gaussian multiaccess channel with an external eavesdropper, and derive a lower bound on the allowed total SDoF under the assumption of perfect instantaneous CSI available at the transmitter and receivers. In \cite{YKPS11}, Yang \textit{et al.} study secure transmission over a two-user MIMO BC with delayed CSIT. They provide an exact characterization of the SDoF region. The coding scheme of \cite{YKPS11} can be seen as an appropriate extension of Maddah Ali-Tse scheme \cite{M-AT12} to accommodate additional noise injection that accounts for security constraints. 

\begin{figure}
\psfragscanon
\begin{center}
\psfrag{W}[][l][.8]{{\tc{blue}{$W_{11}$}},$W_{12}$}
\psfrag{X}[][l][.8]{{\tc{blue}{$W_{21}$}},$W_{22}$}
\psfrag{Y}[][r][.8]{\:\:\:\:\:\:{\tc{blue}{$\hat{W}_{11},\hat{W}_{21}$}}}
\psfrag{Z}[][r][.8]{\:\:\:\:\:\:{$\hat{W}_{12},\hat{W}_{22}$}}
\psfrag{E}[c][c][.7]{\tc{red}{$W_{12}$}}
\psfrag{F}[c][c][.7]{\tc{red}{$W_{22}$}}
\psfrag{G}[c][c][.7]{\tc{red}{$W_{11}$}}
\psfrag{I}[c][c][.7]{\tc{red}{$W_{21}$}}
\psfrag{N}[c][c]{\tc{CadetBlue}{\tf{Tx}$_1$}}
\psfrag{O}[c][c]{\tc{NavyBlue}{\tf{Tx}$_2$}}
\psfrag{P}[c][c]{\tc{CadetBlue}{\tf{Rx}$_1$}}
\psfrag{S}[c][c]{\tc{NavyBlue}{\tf{Rx}$_2$}}
\psfrag{A}[c][c]{$M$}
\psfrag{B}[c][c]{$M$}
\psfrag{C}[c][c]{$N$}
\psfrag{D}[c][c]{$N$}
\psfrag{H}[c][c]{${{\tf{H}}}$}
\psfrag{Q}[c][c]{$({{\tf{y}}}_1^{n-1},{{\tf{H}}}^{n-1})$}
\psfrag{R}[c][c]{$({{\tf{y}}}_2^{n-1},{{\tf{H}}}^{n-1})$}
\psfrag{J}[c][c]{\:\:${{\tf{x}}}_1$}
\psfrag{K}[c][c]{\:\:${{\tf{y}}}_1$}
\psfrag{L}[c][c]{\:\:${{\tf{x}}}_2$}
\psfrag{M}[c][c]{\hspace{-.2em}\:\:${{\tf{y}}}_2$}
\includegraphics[width=0.7\linewidth]{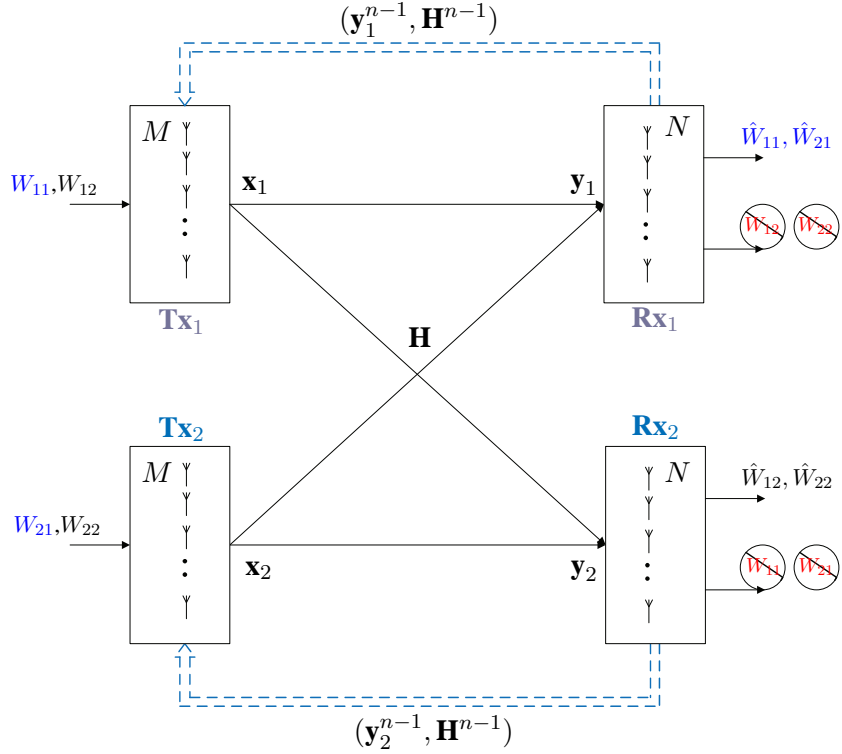}
\end{center}
\caption{MIMO X-channel with asymmetric output feedback and delayed CSIT, with security constraints.}
\psfragscanoff
\label{model}
\end{figure}

In this paper, we consider a two-user MIMO X-channel in which each transmitter is equipped with $M$ antennas, and each receiver is equipped with $N$ antennas as shown in Figure~\ref{model}. Transmitter 1 wants to transmit messages $W_{11}$ and $W_{12}$ to Receiver 1 and Receiver 2, respectively. Similarly, Transmitter 2 wants to transmit messages $W_{21}$ and $W_{22}$ to Receiver 1 and Receiver 2, respectively. The transmission is subject to fast fading effects. Also, we make two assumptions, namely 1) each receiver is assumed to have perfect instantaneous knowledge of its channel coefficients (i.e., CSIR) as well as knowledge of the other receiver's channel coefficients with one unit delay, and 2) there is a noiseless output and CSI feedback from Receiver $i$, $i=1,2$, to Transmitter $i$. We will refer to such output feedback as being \textit{asymmetric}, by opposition to \textit{symmetric} feedback which corresponds to each receiver feeding back its output to \textit{both} transmitters. The considered model is shown in Figure~\ref{model}. Furthermore, the messages that are destined to each receiver are meant to be kept secret from the other receiver. That is, Receiver 2 wants to capture the pair $(W_{11},W_{21})$ of messages that are intended for Receiver 1; and so, in addition to that it is a legitimate receiver of the pair $(W_{12},W_{22})$, it also acts as an eavesdropper on the MIMO multiaccess channel to Receiver 1. Similarly, Receiver 1 wants to capture the pair $(W_{12},W_{22})$ of messages that are intended for Receiver 2; and so, in addition to that it is a legitimate receiver of the pair $(W_{11},W_{21})$, it also acts as an eavesdropper on the MIMO multiaccess channel to Receiver 2. Both eavesdroppers are assumed to be passive, i.e., they are not allowed to modify the transmission. The model that we study can be seen as being that of \cite{TMPS12a} but with security constraints imposed on the transmitted messages. We concentrate on the case of perfect secrecy, and focus on asymptotic behaviors, captured by the allowed secure degrees of freedom over this network model. 

\subsection{Contributions}\label{secI_subsecA}
The main contributions of this paper can be summarized as follows. First, we characterize the sum SDoF region of the two-user $(M,M,N,N)$--MIMO X-channel with asymmetric output feedback and delayed CSIT shown in Figure~\ref{model}. We show that the sum SDoF region of this model is same as the SDoF region of a two-user MIMO BC with delayed CSIT, $2M$ transmit antennas and $N$ antennas at each receiver. This result shows that, for symmetric antennas configurations, the distributed nature of the transmitters does not cause any loss in terms of sum SDoF. The result also emphasizes the usefulness of asymmetric output feedback when used in conjunction with delayed CSIT in securing the transmission of messages in MIMO X-channels, by opposition to in MIMO broadcast channels. That is, for the two-user MIMO X-channel, not only asymmetric output feedback with delayed CSIT does increase the DoF region as shown in \cite{TMPS12a}, it also increases the \textit{secure} DoF region of this network model. The coding scheme that we use for the proof of the direct part is based on an appropriate extension of the one developed by Yang \textit{et al.} \cite{YKPS11} in the context of secure transmission over a two-user MIMO BC with delayed CSIT; and it demonstrates how each transmitter exploits optimally the available output feedback and delayed CSIT.  

Next, concentrating on the role of output feedback in the absence of CSIT from a secrecy degrees of freedom viewpoint, we study two variations of the model of Figure~\ref{model}. In the first model, the transmitters are completely ignorant of the CSI, but are provided with \textit{symmetric} output feedback. As we mentioned previously, this output feedback is assumed to be provided noiselessly by both receivers to both transmitters. In the second model, the transmitters are provided with only asymmetric output feedback, i.e., the model of Figure~\ref{model} but with no CSIT at all.  

For the model with symmetric output feedback at the transmitters, we show that the sum SDoF region is same as the sum SDoF region of the model with asymmetric output feedback and delayed CSIT, i.e., the model of Figure~\ref{model}. In other words, the lack of CSIT does not cause any loss in terms of sum SDoF region as long as each transmitter is provided with output feedback from both recievers. In this case, each transmitter readily gets the side information or interference that is available at the unintended receiver by means of the output feedback; and, therefore, it can align it with the information that is destined to the intended receiver directly, with no need of any CSIT. 

For the model in which only asymmetric output feedback is provided to the transmitters, we establish an inner bound on the sum SDoF region. This inner bound is in general strictly smaller than that of the model of  Figure~\ref{model}; and, so, although its optimality is shown only in some specific cases, it gives insights about the loss incurred by the lack of delayed CSIT. This loss is caused by the fact that, unlike the coding schemes that we develop for the setting with asymmetric output feedback and delayed CSIT and that with symmetric output feedback, for the model with only asymmetric output feedback each transmitter can not learn the side information that is available at the unintended receiver and which is pivotal for the alignment of the interferences in such models. 

Furthermore, we specialize our results to the case in which there are no security constraints. Similar to the setting with security constraints, we show that the optimal sum DoF region of the $(M,M,N,N)$--MIMO X-channel with asymmetric output feedback and delayed CSIT is same of the DoF region of a two-user MIMO BC with $2M$ transmit-antennas, $N$ antennas at each receiver, and delayed CSIT. Finally, we illustrate our results with some numerical examples.

This paper is structured as follows. Section~\ref{secII} provides a formal description of the channel model that we consider, together with some useful definitions. Section~\ref{secIII} states the sum SDoF region of the two-user $(M,M,N,N)$--MIMO X-channel with asymmetric output feedback and delayed CSIT of Figure~\ref{model}. In section~\ref{secIV}, we provide the formal proof of the coding scheme that we use to establish the achievability result. In section~\ref{secV}, we study the role of output feedback in the absence of CSIT. In Section~\ref{secVI}, we specialize the results to the setting with no security constraints; and, in Section~\ref{secVII}, we illustrate our results through some numerical examples. Finally, section~\ref{secVIII} concludes the paper by summarizing its contributions.

\subsection{Notation}

We use the following notations throughout the paper. Boldface upper case letters, e.g., $\dv X$, denote matrices; boldface lower case letters, e.g., $\dv x$, denote vectors; and calligraphic letters designate alphabets, i.e., $\mc X$.  For integers $i \leq j$, we use the notation $\dv X^{j}_{i}$ as a shorthand for $(\dv X_i,\hdots,\dv X_{j})$. The notation $\text{diag}(\{{\tf{H}}[t]\}_t)$ denotes the block diagonal matrix with ${\tf{H}}[t]$ as diagonal elements for all $t$. The Gaussian distribution with mean $\mu$ and variance $\sigma^2$ is denoted by $\mc {CN}(\mu,\sigma^2)$. Finally, throughout the paper, logarithms are taken to base $2$, and the complement to unity of a scalar $u \in [0,1]$ is denoted by $\bar{u}$, i.e., $\bar{u}=1-u$.

\section{System Model and Definitions}\label{secII}

We consider a two-user $(M, M, N, N)$ X-channel, as shown in Figure~\ref{model}. There are two transmitters and two receivers. Both transmitters send messages to both receivers. Transmitter 1 wants to transmit message $W_{11} \in \mc{W}_{11}=\{1,\hdots,2^{nR_{11}(P)} \}$ to Receiver 1, and message $W_{12} \in \mc{W}_{12}=\{1,\hdots,2^{nR_{12}(P)} \}$ to Receiver 2. Similarly, Transmitter 2 wants to transmit message $W_{21} \in \mc{W}_{21}=\{1,\hdots,2^{nR_{21}(P)} \}$ to Receiver 1, and message $W_{22} \in \mc{W}_{22}=\{1,\hdots,2^{nR_{22}(P)} \}$ to Receiver 2. The messages pair $(W_{11},W_{21})$ that is intended to Receiver 1 is meant to be concealed from Receiver 2; and the messages pair $(W_{12},W_{22})$ that is intended to Receiver 2 is meant to be concealed from Receiver 1. Both eavesdroppers are allowed to only overhear the transmission and not modify it, i.e., are assumed to be \textit{passive}.

We consider a fast fading model, and assume that each receiver knows the perfect instantaneous CSI along with the past CSI of the other receiver. Also, we assume that Receiver $i$, $i=1,2$, feeds back its channel output along with the delayed CSI to Transmitter $i$. The outputs received at Receiver 1 and Receiver 2 at each time instant are given by
\begin{align}
\label{gchan_X}
{\tf{{y}}}_1[t] &= {{\tf{H}}}_{11}[t] {{\tf{x}}}_{1}[t]+ {{\tf{H}}}_{12}[t]
{{\tf{x}}}_{2}[t]+{{\tf{z}}}_{1 }[t] \notag\\
{{\tf{y}}}_2[t] &= {{\tf{H}}}_{21}[t] {{\tf{x}}}_{1}[t]+ {{\tf{H}}}_{22}[t]
{{\tf{x}}}_{2}[t]+{{\tf{z}}}_{2 }[t], \:\: t=1,\hdots,n
\end{align}
where ${{\tf{x}}}_i \in \mb{C}^{M}$ is the input vector from Transmitter $i$, $i=1,2$, and ${{\tf{H}}}_{ji} \in \mb{C}^{N \times M}$ is the channel matrix connecting Transmitter $i$ to Receiver $j$, $j=1,2$. We assume arbitrary stationary fading processes, such that ${{\tf{H}}}_{11}[t]$, ${{\tf{H}}}_{12}[t]$, ${{\tf{H}}}_{21}[t]$ and ${{\tf{H}}}_{22}[t]$ are mutually independent and change independently across time. The noise vectors ${{\tf{z}}}_{j}[t] \in \mb{C}^{N}$ are assumed to be independent and identically distributed (i.i.d.) white Gaussian, with ${{\tf{z}}}_j \sim \mc{CN}({\bf{0}},{{\tf{I}}}_{N})$ for $j=1,2$. Furthermore, we consider average block power constraints on the transmitters inputs, as 
\begin{align}
\label{p_con1}
&\sum_{t=1}^n \mathbb{E}[\| {{\tf{x}}}_{i}[t]\|^2] \leq nP, \qquad
\text{for}\:\:\: i \in \{1,2\}.
\end{align}
For convenience, we let $\tf{H}[t]$ = $\left [
\begin{smallmatrix}
\tf{H}_{11}[t] & {{\tf{H}}}_{12}[t]\\
\tf{H}_{21}[t] & {{\tf{H}}}_{22}[t]
\end{smallmatrix} \right]$ designate the channel state matrix and ${{\tf{H}}}^{t-1} =\{{{\tf{H}}}[1],\hdots,{{\tf{H}}}[t-1]\}$ designate the collection of channel state matrices for the past $(t-1)$ symbols. For convenience, we set ${{\tf{H}}}^{0} =\emptyset$. We assume that, at each time instant $t$, the channel state matrix ${{\tf{H}}}[t]$ is full rank almost surely. Also, we denote by ${{\tf{y}}}_j^{t-1} =\{{{\tf{y}}}_j[1],\hdots,{{\tf{y}}}_j[t-1]\}$ the collection of the outputs at Receiver $j$, $j=1,2$, over the past $(t-1)$ symbols. At each time instant $t$, the past states of the channel $\tf{H}^{t-1}$ are known to all terminals. However the instantaneous states $(\tf{H}_{11}[t],\tf{H}_{12}[t])$ are known only to Receiver 1, and the instantaneous states $(\tf{H}_{21}[t],\tf{H}_{22}[t])$ are known only to Receiver 2. Furthermore, at each time instant, Receiver 1 feeds back the output vector ${{\tf{y}}}_1^{t-1}$ to Transmitter 1, and Receiver 2 feeds back the output vector ${{\tf{y}}}_2^{t-1}$ to Transmitter 2.

From a practical viewpoint, the two-user MIMO X-channel with asymmetric output feedback and delayed CSIT of Figure~\ref{model} may model a cellular network in which two base stations communicate with two destinations. Each base station sends information messages to both receivers; and, in doing so, it wants to keep the information that is sent to each receiver secret from the other receiver. Here, by opposition to classic wiretap channels in which the eavesdropper is generally not willing to feed back information about its channel to the transmitter from which it wants to intercept the transmission, each receiver is not merely an eavesdropper for the information sent by the transmitters to the other receiver but is also a legitimate receiver intended to get other information messages from the \textit{same} transmitters. For this reason, in its desire to help the transmitters obtain a better estimate of the channel, the receivers may find it useful to feedback information on their channels to the transmitters. Depending on the strength of the feedback signal, this may be heard at both or only one of the transmitters.

\vspace{.5em}
\begin{definition}\label{definition1}
A code for the Gaussian $(M,M,N,N)$--MIMO X-channel with asymmetric output feedback and delayed CSIT consists of two sequences of stochastic encoders at the transmitters,
\begin{align}
\{\phi_{1t} \:\: &: \:\: \mc W_{11}{\times}\mc W_{12}{\times}\mc H^{t-1}{\times}\mc Y_1^{N(t-1)} \longrightarrow \mc X_1^M\}_{t=1}^{n}\nonumber\\
\{\phi_{2t} \:\: &: \:\: \mc W_{21}{\times}\mc W_{22}{\times}\mc H^{t-1}{\times}\mc Y_2^{N(t-1)}  \longrightarrow \mc X_2^M\}_{t=1}^{n}
\end{align}
where the messages $W_{11}$, $W_{12}$, $W_{21}$ and $W_{22}$ are drawn uniformly over the sets $\mc W_{11}$, $\mc W_{12}$, $\mc W_{21}$ and $\mc W_{22}$, respectively; and four decoding functions at the receivers,
\begin{align}
\psi_{11} \:\: &: \:\: \mc Y_1^{Nn}{\times}\mc H^{n-1}{\times} \mc H_{11}{\times}\mc H_{12} \longrightarrow \hat{\mc{W}}_{11}\nonumber\\
\psi_{21} \:\: &: \:\: \mc Y_1^{Nn}{\times}\mc H^{n-1}{\times}\mc H_{11}{\times}\mc H_{12} \longrightarrow \hat{\mc{W}}_{21}\nonumber\\
\psi_{12} \:\: &: \:\: \mc Y_2^{Nn}{\times}\mc H^{n-1}{\times}\mc H_{21}{\times}\mc H_{22} \longrightarrow \hat{\mc{W}}_{12}\nonumber\\
\psi_{22} \:\: &: \:\: \mc Y_2^{Nn}{\times}\mc H^{n-1}{\times}\mc H_{21}{\times}\mc H_{22} \longrightarrow \hat{\mc{W}}_{22}.
\end{align}
\end{definition}
\vspace{.5em}
\begin{definition}\label{definition2}
A rate quadruple $(R_{11}(P),R_{12}(P),R_{21}(P),R_{22}(P))$ is said to be achievable if there exists a sequence of codes such that,
\begin{equation}
 \limsup_{n \rightarrow \infty} \text{Pr}\{\hat{W}_{ij} \neq W_{ij} | W_{ij}\}=0, \: \forall\: (i,j) \in \{1,2\}^2.
\end{equation}
\end{definition}
\vspace{.5em}
\begin{definition}\label{definition3}
A SDoF quadruple $(d_{11},d_{12},d_{21},d_{22})$ is said to be achievable if there exists a sequence of codes satisfying the following reliability conditions at both receivers,
\begin{align}
&\lim_{P \rightarrow \infty} \liminf_{n \rightarrow \infty} \frac{\log |\mc W_{ij}(n,P)|}{n\log P} \geq d_{ij}, \quad \forall\:\: (i,j) \in \{1,2\}^2 \nonumber\\
& \limsup_{n \rightarrow \infty} \text{Pr}\{\hat{W}_{ij} \neq W_{ij} | W_{ij}\}=0,\: \forall (i,j) \in \{1,2\}^2
\end{align}
as well as the perfect secrecy conditions
\begin{align}
& \lim_{P \rightarrow \infty} \limsup_{n \rightarrow \infty} \frac{I(W_{12},W_{22};{{\tf{y}}}_1^n,{{\tf{H}}}^n)}{n\log P}=0 \nonumber\\
& \lim_{P \rightarrow \infty} \limsup_{n \rightarrow \infty} \frac{I(W_{11},W_{21};{{\tf{y}}}_2^n,{{\tf{H}}}^n)}{n\log P}=0.
\end{align}
\end{definition}

\begin{definition}\label{definition4}
We define the sum secure degrees of freedom region of the MIMO X-channel with asymmetric output feedback and delayed CSIT, which we denote by $\mc C^{\text{sum}}_{\text{SDoF}}$, as the set of all of all pairs $(d_{11}+d_{21}, d_{12}+d_{22})$ for all achievable non-negative quadruples $(d_{11},d_{21},d_{12},d_{22})$. We also define the total secure degrees of freedom as $\text{SDoF}^{\text{d-CSIT,F}}_{\text{total}}=\max_{(d_{11},d_{21},d_{12},d_{22})} d_{11}+d_{21}+d_{12}+d_{22}$. 
\end{definition}

\section{Sum SDoF of $(M,M,N,N)$--MIMO X-channel with asymmetric output feedback and delayed CSIT}\label{secIII}

In this section we state our main result on the optimal sum SDoF region of the two-user MIMO X-channel with asymmetric output feedback and delayed CSIT. We illustrate our result by providing few examples which give insights into the proposed coding scheme. 

\noindent For convenience we define the following quantity that we will use extensively in the sequel. Let, for given non-negative $(M,N)$,
\begin{equation}
d_s(N,N,M)=\left\{
\begin{array}{ll}
0 & \:\:\: \text{if} \:\:\: M \leq N\\
\frac{NM(M-N)}{N^2+M(M-N)} & \:\:\:\text{if} \:\:\: N \leq M \leq 2N\\
\frac{2N}{3} &\:\:\: \text{if} \:\:\: M \geq 2N
\end{array}
\right.
\end{equation}
\noindent The following theorem characterizes the sum SDoF region of the MIMO X-channel with asymmetric output feedback and delayed CSIT.

\vspace{.5em}

\begin{theorem}\label{theorem-sum-sdof-region-mimo-x-channel-with-asymmetric-output-feedback-and-delayed-csi}
The sum SDoF region $\mc C^{\text{sum}}_{\text{SDoF}}$ of the two-user $(M,M,N,N)$--MIMO X-channel with asymmetric output feedback and delayed CSIT is given by the set of all non-negative pairs $(d_{11}+d_{21},d_{12}+d_{22})$ satisfying
\begin{align}
\frac{d_{11}+d_{21}}{d_s(N,N,2M)}+\frac{d_{12}+d_{22}}{\min(2M,2N)} &\leq 1 \nonumber\\
\frac{d_{11}+d_{21}}{\min(2M,2N)}+\frac{d_{12}+d_{22}}{d_s(N,N,2M)} &\leq 1
\label{linear-equations-corner-points-theorem-sum-sdof-region-mimo-x-channel-with-asymmetric-output-feedback-and-delayed-csi}
\end{align}
for $2M \geq N$; and $\mc C^{\text{sum}}_{\text{SDoF}}=\{(0,0)\}$ if $2M \leq N$.
\end{theorem}

\vspace{.5em}

\begin{IEEEproof}
The converse proof follows by allowing the transmitters to cooperate and then using the outer bound established in \cite[Theorem 3]{YKPS11} in the context of secure transmission over MIMO broadcast channels with delayed CSIT, by taking $2M$ transmit antennas and $N$ antennas at each receiver. Note that Theorem 3 of \cite{YKPS11} continues to hold if one provides additional feedback from the receivers to the transmitter. The proof of achievability is given in Section~\ref{secIV}.
\end{IEEEproof}

\vspace{.5em}

\begin{remark}\label{remark-corner-points-theorem-sum-sdof-region-mimo-x-channel-with-asymmetric-output-feedback-and-delayed-csi}
In the case in which $2M \geq N$, the sum SDoF region of Theorem~\ref{theorem-sum-sdof-region-mimo-x-channel-with-asymmetric-output-feedback-and-delayed-csi} is characterized fully  by the three corner points $(d_s(N,N,2M),0)$, $(0,d_s(N,N,2M))$ and
\begin{align}
&(d_{11}+d_{21},d_{12}+d_{22}) = \nonumber\\
&\quad \left\{
\begin{array}{ll}
\big(\frac{N(2M-N)}{2M},\frac{N(2M-N)}{2M}\big) & \:\:\:\text{if}\:\:\: N \leq 2M \leq 2N\\
\big(\frac{N}{2},\frac{N}{2}\big) & \:\:\:\text{if}\:\:\:  2N \leq 2M
\end{array}
\right.
\label{intersection-corner-point-theorem-sum-sdof-region-mimo-x-channel-with-asymmetric-output-feedback-and-delayed-csi}
\end{align}
\end{remark}

\vspace{.5em}

\begin{remark}\label{remark-equivalence-mimo-x-channel-with-asymmetric-output-feedback-and-delayed-csi-bc-with-delayed-csi}
The sum SDoF region of Theorem~\ref{theorem-sum-sdof-region-mimo-x-channel-with-asymmetric-output-feedback-and-delayed-csi} is same as the SDoF region of a two-user MIMO BC with delayed CSIT in which the transmitter is equipped with $2M$ antennas and each receiver is equipped with $N$ antennas\cite[Theorem 3]{YKPS11}. Therefore, Theorem~\ref{theorem-sum-sdof-region-mimo-x-channel-with-asymmetric-output-feedback-and-delayed-csi} shows that there is no performance loss in terms of total SDoF due to the distributed nature of the transmitters in the MIMO X-channel that we consider. Note that, in particular, this implies that, like the setting with no security constraints~\cite[Theorem 1]{TMPS12a}, the total secure degrees of freedom, defined as in Definition~\ref{definition4} and given by
\begin{equation}
\text{SDoF}^{\text{d-CSIT,F}}_{\text{total}}=\left\{
\begin{array}{ll}
0  & \:\:\:\text{if}\:\:\: 2M \leq N\\
\frac{N(2M-N)}{M} & \:\:\:\text{if}\:\:\: N \leq 2M \leq N\\
N & \:\:\:\text{if}\:\:\: 2M \geq 2N
\end{array}
\right.
\label{total-sdof-mimo-x-channel-with-asymmetric-output-feedback-and-delayed-csi-bc-with-delayed-csi}
\end{equation}
is also preserved upon the availability of asymmetric output feedback and delayed CSI at the transmitters, even though the transmitters are distributed.
\end{remark}

\vspace{.5em}

\begin{figure}
\psfragscanon
\begin{center}
\psfrag{data1}[l][l][.45]{\hspace{-.5em}{$(M,N) = (4,4),\:\: 2M \geq 2N$}}
\psfrag{data2}[l][l][.45]{\hspace{-.5em}{$(M,N)= (2,3),\:\: N \leq 2M \leq 2N$}}
\psfrag{data3}[l][l][.45]{\hspace{0cm}{$2M \leq N$}}
\psfrag{y}[c][c][.75]{$d_{12}+d_{22}$}
\psfrag{x}[c][c][.75]{$d_{11}+d_{21}$}
\psfrag{n}[c][c][.75]{\hspace{.5em}$(\frac{3}{4},\frac{3}{4})$}
\psfrag{m}[c][c][.75]{\hspace{.5em}$(2,2)$}
\includegraphics[width=0.8\linewidth]{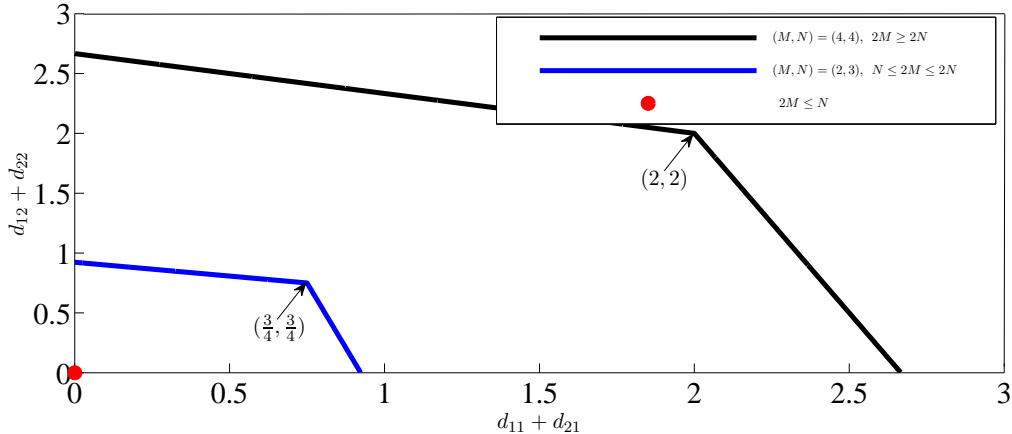}
\end{center}
\caption{Sum SDoF region of the $(M,M,N,N)$--MIMO X-channel with asymmetric output feedback and delayed CSIT, for different antennas configurations.}
\psfragscanoff
 \label{fig-illustration-sum-sdof}
\end{figure}

\begin{table}
\begin{center}
\begin{tabular}{| c | c | c | c |}
\hline Case & $\text{SDoF}^{\text{d-CSIT,F}}_{\text{total}}$ &
$\text{DoF}^{\text{d-CSIT,F}}_{\text{total}}$ \cite{TMPS12a} &
$\text{DoF}^{\text{n-CSIT,nF}}_{\text{total}}$ \cite{vaze_no} 
\\ [.5ex] \hline
$2M \leq N$ & $0$ & $2M$ & $2M$ \\ [.5ex] \hline
$N \leq 2M \leq 2N$ & $\frac{N(2M-N)}{M}$ & $\frac{4MN}{2M+N}$ & $N$  \\ [.5ex] \hline
$2N \leq 2M $ & $N$ & $\frac{4N}{3}$ & $N$  \\ [.5ex]\hline
\end{tabular}
\vspace{1em}
\caption{Total SDoF and total DoF of $(M,M,N,N)$--MIMO X-channels with different degrees of output feedback and delayed CSIT.}
\end{center}
\label{table1}
\end{table}

\begin{figure}
\psfragscanon
\begin{center}
\psfrag{data1}[l][l][.5]{\hspace{0em}{Total secure DoF with asymmetric feedback and delayed CSIT~\eqref{total-sdof-mimo-x-channel-with-asymmetric-output-feedback-and-delayed-csi-bc-with-delayed-csi}}}
\psfrag{data2}[l][l][.45]{\hspace{0em}{Total DoF with asymmetric feedback and delayed CSIT \cite[Theorem~1]{TMPS12a}}}
\psfrag{data3}[l][l][.5]{\hspace{0em}{Total DoF with no feedback and no CSIT \cite[Theorem~11]{vaze_no}}}
\psfrag{y}[c][c][.75]{$d_{11}+d_{12}+d_{21}+d_{22}$}
\psfrag{x}[c][c][.75]{Number of transmit antennas $M$ at each transmitter}
\includegraphics[width=0.8\linewidth]{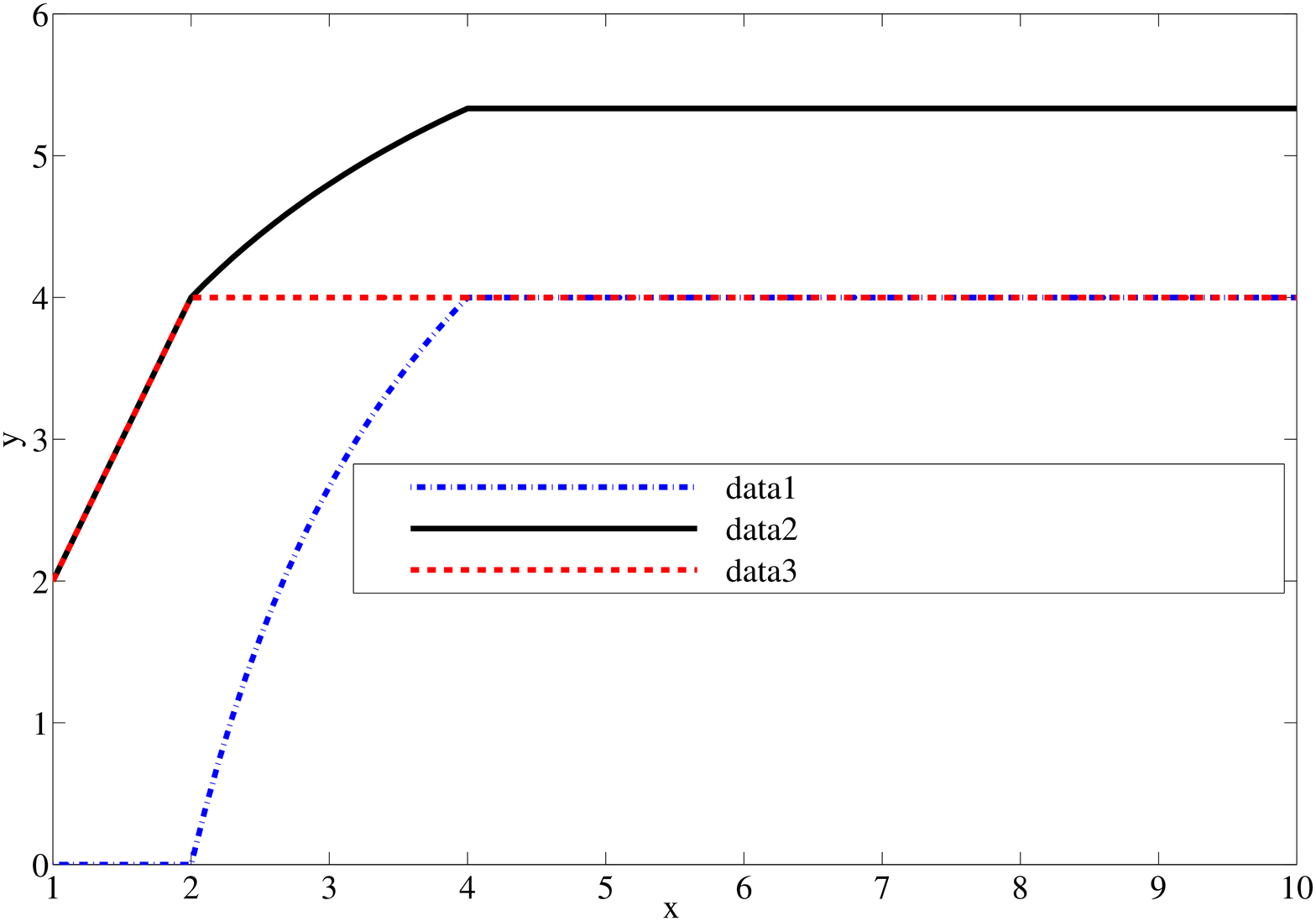}
\end{center}
\caption{Total secure degrees of freedom of the $(M,M,N,N)$--MIMO X-channel as a function of the  number of transmit antennas $M$ at each transmitter, for a fixed number $N=4$ of receive antennas at each receiver.}
\psfragscanoff
 \label{fig-illustration-total-sdof}
\end{figure}

\vspace{.5em}


Figure~\ref{fig-illustration-sum-sdof} illustrates the optimal sum SDoF region of the  $(M,M,N,N)$--MIMO X-channel with asymmetric output feedback and delayed CSIT as given in Theorem~\ref{theorem-sum-sdof-region-mimo-x-channel-with-asymmetric-output-feedback-and-delayed-csi}, for different  values of the transmit- and receive-antennas. Obviously, secure messages transmission is not possible if, accounting for the antennas available at both transmitters, there are less transmit antennas than receive antennas at each receiver, i.e., $2M \leq N$. Also, the sum SDoF region increases with the pair $(M,N)$ if $N \leq 2M \leq 2N$. For a given number $N$ of receiver antennas at each receiver, the sum SDoF region no longer increases with the number of transmit-antennas $M$ at each transmitter as long as $M \geq N$. This shows that, from a SDoF perspective, there is no gain from equipping the transmitters with more than $N$ antennas each. A similar behavior is shown in Table I and Figure~\ref{fig-illustration-total-sdof} from a total secure degrees of freedom viewpoint. Table I summarizes the optimal total SDoF of the $(M,M,N,N)$--MIMO X-channel with asymmetric output feedback and delayed CSIT as given by ~\eqref{total-sdof-mimo-x-channel-with-asymmetric-output-feedback-and-delayed-csi-bc-with-delayed-csi}, as well as the total DoF of the  $(M,M,N,N)$--MIMO X-channel without security constraints, with asymmetric output feedback and delayed CSIT~\cite[Theorem 1]{TMPS12a} and with no output feedback and no CSIT~\cite[Theorem 11]{vaze_no}. Figure~\ref{fig-illustration-total-sdof} depicts the evolution of the total SDoF~\eqref{total-sdof-mimo-x-channel-with-asymmetric-output-feedback-and-delayed-csi-bc-with-delayed-csi} as a function of the number of transmit antennas at each transmitter, for an example configuration in which each receiver is equipped with $N=4$ antennas. It is interesting to note that, for the case $M \geq N$ the total SDoF of the MIMO X-channel with asymmetric output feedback and  delayed CSIT is same as the DoF of the MIMO X-channel with no feedback and no CSIT. Thus, providing the transmitters with asymmetric output feedback and delayed CSIT can be interpreted as the price for secrecy in this case.

\section{Proof of Direct Part of Theorem~\ref{theorem-sum-sdof-region-mimo-x-channel-with-asymmetric-output-feedback-and-delayed-csi}}\label{secIV}

In this section, we provide a description of the coding scheme that we use for the proof of Theorem~\ref {theorem-sum-sdof-region-mimo-x-channel-with-asymmetric-output-feedback-and-delayed-csi}. This coding scheme can be seen as an extension, to the case of non-cooperative or distributed transmitters, of that established by Yang \textit{et al.} \cite{YKPS11} in the context of secure transmission over a two-user MIMO BC with delayed CSIT. 

\noindent In the case in which $2M \leq N$, every receiver has enough antennas to decode all of the information that is sent by the transmitters; and, so,  secure transmission of messages is not possible. In the case in which $2M \geq N$, it is enough to prove that the corner points that are given in Remark~\ref{remark-corner-points-theorem-sum-sdof-region-mimo-x-channel-with-asymmetric-output-feedback-and-delayed-csi} are achievable, since the entire region can then be achieved by time-sharing. The achievability of each of the two corner points $(d_s(N,N,2M),0)$ follows by the coding scheme of~\cite[Theorem 1]{YKPS11}, by having the transmitters sending information messages only to one receiver and the other receiver acting as an eavesdropper. In what follows, we show that the point given by \eqref{intersection-corner-point-theorem-sum-sdof-region-mimo-x-channel-with-asymmetric-output-feedback-and-delayed-csi} is achievable. We divide the analysis into two cases.

\subsection{Case 1: $N \leq 2M \leq 2N$}\label{secIV_subsecA}

The achievability in this case follows by a careful combination of Maddah Ali-Tse coding scheme~\cite{M-AT12} developed for the MIMO broadcast channel with additional noise injection. Also, as we already mentioned, it has connections with, and can be seen as an extension to the case of distributed transmitters, of that developed by Yang {\it{et al.}} \cite{YKPS11} in the context of secure transmission over a two-user MIMO BC with delayed CSIT. The scheme also extends Tandon \textit{et al.} \cite{TMPS12a} coding scheme about X-channels without security constraints to the setting with secrecy. The communication takes place in four phases. For simplicity of the analysis, and in accordance with the DoF framework, we ignore the additive noise impairment.
\vspace{.5em}
 
\noindent\textit{\textbf{Phase 1: Injecting artificial noise}}

\noindent In the first phase, the communication takes place in $T_1= N^2$ channel uses. Let ${\tf{{u}}}_{1}  = [u^1_{1},\hdots,u^{MT_1}_{1}]^T$ and ${\tf{{u}}}_{2} = [u^1_{2},\hdots,u^{MT_1}_{2}]^T$ denote the artificial noises injected by Transmitter 1 and Transmitter 2 respectively.
The channel outputs at Receiver 1 and  Receiver 2 during this phase are given by
\begin{eqnarray}
{\bf{y}}_{1}^{(1)} = \tilde{\bf{H}}_{11}^{(1)} {\tf{u}}_1+ \tilde{\bf{H}}_{12}^{(1)} {\tf{u}}_2 \\
{\bf{y}}_{2}^{(1)} = \tilde{\bf{H}}_{21}^{(1)} {\tf{u}}_1+ \tilde{\bf{H}}_{22}^{(1)} {\tf{u}}_2
\end{eqnarray}
where $\tilde{{\tf{H}}}_{ji}^{(1)} = \text{diag}(\{{{\tf{H}}}_{ji}^{(1)}[t]\}_t) \in \mb{C}^{NT_1 \times MT_1}$, for $t = 1,\hdots,T_1$, $i=1,2$, $j = 1,2$, ${\bf{y}}_{1}^{(1)} \in \mb{C}^{NT_1}$ and ${\bf{y}}_{2}^{(1)} \in \mb{C}^{NT_1}$. During this phase, each receiver gets $NT_1$ linearly independent equations that relate $2MT_1$ ${\tf{{u}}}_{1}$- and ${\tf{{u}}}_{2}$-variables. At the end of this phase, the channel output at Receiver $i$, $i=1,2$, is fed back along with the past CSI to Transmitter $i$. 
\vspace{.5em}

\noindent\textit{\textbf{Phase 2: Fresh information for Receiver 1}}

\noindent In this phase, the communication takes place in $T_2=N(2M-N)$ channel uses. Both transmitters transmit to Receiver 1 confidential messages that they want to conceal from Receiver 2. To this end, Transmitter 1 sends fresh information ${\tf{v}}_{11} = [v^1_{11},\hdots,v^{MT_2}_{11}]^T$ along with a linear combination of the channel output ${\bf{y}}_{1}^{(1)}$ of Receiver 1 during the first phase; and Transmitter 2 sends only fresh information ${\tf{{v}}}_{21} =[v^1_{21},\hdots,v^{MT_2}_{21}]^T$ intended for Receiver 1, i.e.,
\begin{align}
{\tf{x}}_1 &= {\tf{v}}_{11}+ \Theta_1{\bf{y}}_{1}^{(1)} \nonumber\\
{\tf{x}}_2 &= {\tf{v}}_{21}
\label{inputs-phase2-proof-of-theorem1}
\end{align}
where $\Theta_1 \in \mc{C}^{MT_2 \times NT_1}$ is a matrix that is known at all nodes and whose choice will be specified below. The channel outputs at the receivers during this phase are given by
\begin{subequations}
\begin{align}
\label{output-receiver1-phase2-proof-of-theorem1}
{\bf{y}}_{1}^{(2)} &= \tilde{\bf{H}}_{11}^{(2)} ({\tf{v}}_{11}+ \Theta_1 {\bf{y}}_{1}^{(1)})+ \tilde{\bf{H}}_{12}^{(2)} {\tf{v}}_{21} \\
{\bf{y}}_{2}^{(2)} &= \tilde{\bf{H}}_{21}^{(2)} ({\tf{v}}_{11}+ \Theta_1 {\bf{y}}_{1}^{(1)})+ \tilde{\bf{H}}_{22}^{(2)} {\tf{v}}_{21} 
\label{output-receiver2-phase2-proof-of-theorem1}
\end{align}
\label{outputs-receivers-phase2-proof-of-theorem1}
\end{subequations}

\noindent where $\tilde{{\tf{H}}}_{ji}^{(2)} = \text{diag}(\{{{\tf{H}}}_{ji}^{(2)}[t]\}_t) \in \mb{C}^{NT_2 \times MT_2}$, for $t = 1,\hdots,T_2$, $i=1,2$, $j = 1,2$, ${\bf{y}}_{1}^{(2)} \in \mb{C}^{NT_2}$  and ${\bf{y}}_{2}^{(2)} \in \mb{C}^{NT_2}$. At the end of this phase, the channel output at Receiver $i$, $i=1,2$, is fed back along with the delayed CSI to Transmitter $i$. 

\noindent Since Receiver 1 knows the CSI $(\tilde{\bf{H}}_{11}^{(2)},\tilde{\bf{H}}_{12}^{(2)})$ and the channel output ${\tf{{y}}}^{(1)}_1$ from Phase 1, it subtracts out the contribution of  ${\tf{{y}}}^{(1)}_1$ from the received signal ${\bf{y}}_{1}^{(2)}$ and, thus, obtains $NT_2$ linearly independent equations with $2MT_2$ ${\tf{{v}}}_{11}$- and ${\tf{{v}}}_{21}$-variables. Thus, Receiver 1 requires $(2M-N)T_2$ extra linearly independent equations to successfully decode the ${\tf{{v}}}_{11}$- and ${\tf{{v}}}_{21}$-symbols that are intended to it during this phase. Let $\tilde{\bf{y}}_{2}^{(2)} \in \mb{C}^{(2M-N)T_2}$ denote a set of $(2M-N)T_2$ such linearly independent equations, selected among the available $NT_2$ side information equations ${\bf{y}}_{2}^{(2)} \in \mb{C}^{NT_2}$ (recall that $2M-N \leq N$ in this case). If these equations can be conveyed to Receiver 1, they will suffice to help it decode the ${\tf{{v}}}_{11}$- and ${\tf{{v}}}_{21}$-symbols, since the latter already knows ${\tf{{y}}}^{(1)}_1$. These equations will be transmitted \textit{jointly} by the two transmitters in Phase 4, and are learned as follows. Transmitter 2 learns ${\tf{{y}}}^{(2)}_2$, and so $\tilde{\bf{y}}_{2}^{(2)}$, directly by means of the output feedback from Receiver 2 at the end of this phase. Transmitter 1 learns  ${\tf{{y}}}^{(2)}_2$, and so $\tilde{\bf{y}}_{2}^{(2)}$, by means of output as well as delayed CSI feedback from Receiver 1 at the end of Phase 2, as follows.  First, Transmitter 1 utilizes the fed back output ${\tf{{y}}}^{(2)}_1$ to learn  the ${\tf{{v}}}_{21}$-symbols that are transmitted by Transmitter 2 during this phase. This can be accomplished correctly since Transmitter 1, which already knows ${\tf{{v}}}_{11}$ and ${\tf{{y}}}^{(1)}_1$, has also gotten the delayed CSI $(\tilde{\bf{H}}_{11}^{(2)},\tilde{\bf{H}}_{12}^{(2)})$ and $M \leq N$. Next, Transmitter 1, which also knows the delayed CSI $(\tilde{\bf{H}}_{21}^{(2)},\tilde{\bf{H}}_{22}^{(2)})$, reconstructs ${\tf{{y}}}^{(2)}_2$ as given by \eqref{output-receiver2-phase2-proof-of-theorem1}. 
\vspace{.5em}

\noindent\textit{\textbf{Phase 3: Fresh information for Receiver 2}}

\noindent This phase is similar to Phase 2, with the roles of Transmitter 1 and Transmitter 2, as well as those of Receiver 1 and Receiver 2, being swapped. More specifically, the communication takes place in $T_2=N(2M-N)$ channel uses. Fresh information is sent by both transmitters to Receiver 2, and is to be concealed from Receiver 1. Transmitter 1 transmits fresh information ${\tf{{v}}}_{12} = [v^1_{12},\hdots,v^{MT_2}_{12}]^T$ to Receiver 2, and Transmitter 2 transmits ${\tf{{v}}}_{22} = [v^1_{22},\hdots,v^{MT_2}_{22}]^T$ along with a linear combination of the channel output ${\bf{y}}_{2}^{(1)}$ at Receiver 2 during Phase 1, i.e.,
\begin{align}
{\tf{x}}_1 &= {\tf{v}}_{12} \nonumber\\
{\tf{x}}_2 &= {\tf{v}}_{22} + \Theta_2 {\bf{y}}_{2}^{(1)}
\label{inputs-phase3-proof-of-theorem1}
\end{align}
\noindent where $\Theta_2 \in \mc{C}^{MT_2 \times NT_1 }$ is matrix that is known at all nodes and whose choice will be specified below. The channel outputs during this phase are given by
\begin{subequations}
\begin{align}
\label{output-receiver1-phase3-proof-of-theorem1}
{\bf{y}}_{1}^{(3)} &= \tilde{\bf{H}}_{11}^{(3)} {\tf{v}}_{12}+ \tilde{\bf{H}}_{12}^{(3)} ({\tf{v}}_{22}+ \Theta_2 {\bf{y}}_{2}^{(1)}) \\
{\bf{y}}_{2}^{(3)} &= \tilde{\bf{H}}_{21}^{(3)} {\tf{v}}_{12}+ \tilde{\bf{H}}_{22}^{(3)} ({\tf{v}}_{22}+ \Theta_2 {\bf{y}}_{2}^{(1)})
\label{output-receiver2-phase3-proof-of-theorem1}
\end{align}
\label{outputs-receivers-phase3-proof-of-theorem1}
\end{subequations}

\noindent where $\tilde{{\tf{H}}}_{ji}^{(3)} = \text{diag}(\{{{\tf{H}}}_{ji}^{(3)}[t]\}_t) \in \mc{C}^{NT_2 \times MT_2}$ for $t = 1,\hdots,T_2$, $i=1,2$, $j=1,2$, ${\bf{y}}_{1}^{(3)} \in \mb{C}^{NT_2}$  and ${\bf{y}}_{2}^{(3)} \in \mb{C}^{NT_2}$. At the end of this phase, the channel output at Receiver $i$, $i=1,2$, is fed back along with the delayed CSI to Transmitter $i$.

\noindent Similar to Phase 2, at the end of Phase 3 since Receiver 2 knows the CSI $(\tilde{\bf{H}}_{21}^{(3)},\tilde{\bf{H}}_{22}^{(3)})$ and the channel output ${\tf{{y}}}^{(1)}_2$ from Phase 1, it subtracts out the contribution of ${\tf{{y}}}^{(1)}_2$ from the received signal ${\bf{y}}_{2}^{(3)}$ and, thus, obtain $NT_2$ linearly independent equations with $2MT_2$ ${\tf{{v}}}_{12}$- and ${\tf{{v}}}_{22}$-variables. Thus, similar to Receiver 1 at the end of Phase 2, Receiver 2 requires $(2M-N)T_2$ extra linearly independent equations to successfully decode the ${\tf{{v}}}_{12}$- and ${\tf{{v}}}_{22}$-symbols that are intended to it during this phase. Let $\tilde{\bf{y}}_{1}^{(3)} \in \mb{C}^{(2M-N)T_2}$ denote a set of $(2M-N)T_2$ such linearly independent equations, selected among the available $NT_2$ side information equations ${\bf{y}}_{1}^{(3)} \in \mb{C}^{NT_2}$. If these equations can be conveyed to Receiver 2, they will suffice to help it decode the ${\tf{{v}}}_{12}$- and ${\tf{{v}}}_{22}$-symbols, since the latter already knows ${\tf{{y}}}^{(1)}_2$. These equations will be transmitted \textit{jointly} by the two transmitters in Phase 4, and are learned as follows. Transmitter 1 learns ${\tf{{y}}}^{(3)}_1$, and so $\tilde{\bf{y}}_{1}^{(3)}$, directly by means of the output feedback from Receiver 1 at the end of this phase. Transmitter 2 learns  ${\tf{{y}}}^{(3)}_1$, and so $\tilde{\bf{y}}_{1}^{(3)}$, by means of output as well as delayed CSI feedback from Receiver 2 at the end of Phase 3, as follows.  First, Transmitter 2 utilizes the fed back output ${\tf{{y}}}^{(3)}_2$ to learn  the ${\tf{{v}}}_{12}$-symbols that are transmitted by Transmitter 1 during this phase. This can be accomplished correctly since Transmitter 2, which already knows ${\tf{{v}}}_{22}$ and ${\tf{{y}}}^{(1)}_2$, has also gotten the delayed CSI $(\tilde{\bf{H}}_{21}^{(3)},\tilde{\bf{H}}_{22}^{(3)})$ and $M \leq N$. Next, Transmitter 2, which also knows the delayed CSI $(\tilde{\bf{H}}_{11}^{(3)},\tilde{\bf{H}}_{12}^{(3)})$, reconstructs ${\tf{{y}}}^{(3)}_1$ as given by \eqref{output-receiver1-phase3-proof-of-theorem1}.  
\vspace{.5em}

\noindent\textit{\textbf{Phase 4: Interference alignment and decoding}}

\noindent Recall that, at the end of Phase 3, Receiver 1 requires $(2M-N)T_2$ extra equations to successfully decode the sent ${\tf{{v}}}_{11}$- and ${\tf{{v}}}_{21}$-symbols, and Receiver 2 requires $(2M-N)T_2$ extra equations to successfully decode the sent ${\tf{{v}}}_{12}$- and ${\tf{{v}}}_{22}$-symbols. Also, recall that at the end of this third phase, \textit{both} transmitters can re-construct the side information, or interference, equations $\tilde{\bf{y}}_{1}^{(3)} \in \mb{C}^{(2M-N)T_2}$ and $\tilde{\bf{y}}_{2}^{(2)} \in \mb{C}^{(2M-N)T_2}$ that are required by both receivers. In this phase, both transmitters transmit these equations jointly, as follows.  

\noindent The communication takes place in $T_3=(2M-N)^2$ channel uses. Let
\begin{eqnarray*}
I = \Phi_{1}[\underbrace{\tilde{\bf{y}}_{2}^{(2)}}_{(2M-N)T_2} \:\: \underbrace{\phi}_{(2N-2M)T_2} ]^T+\Phi_{2}[\underbrace{\tilde{\bf{y}}_{1}^{(3)}}_{(2M-N)T_2 } \:\: \underbrace{\phi}_{(2N-2M)T_2 }]^T
\end{eqnarray*}
\noindent where $\Phi_{1} \in \mc{C}^{2MT_3 \times NT_2}$ and  $\Phi_{2} \in  \mc{C}^{2MT_3 \times NT_2}$ are linear combination matrices that are assumed to be known to all the nodes.
During this phase, the transmitters send
\begin{eqnarray*}
&&{\tf{{x}}}_1 = [I^1, \hdots, I^{MT_3}] \notag\\
&&{\tf{{x}}}_2 = [I^{(M+1)T_3}, \hdots, I^{2MT_3}].
\end{eqnarray*}
\noindent At the end of Phase 4, Receiver 1 gets $NT_3$ equations in $2NT_3$ variables. Since Receiver 1 knows  ${\bf{y}}_{1}^{(3)}$ from Phase 3 as well as the CSI, it can subtract out the contribution of $\tilde{\bf{y}}_{1}^{(3)}$ from its received signal to get $NT_3$ equations in $NT_3$ variables. Thus, Receiver 1 can recover the $\tilde{\bf{y}}_{2}^{(2)} \in \mb{C}^{(2M-N)T_2}$ interference equations. Then, using the pair of output vectors $({\bf{y}}_{1}^{(2)},\tilde{\bf{y}}_{2}^{(2)})$, Receiver 1 first subtracts out the contribution of ${\bf{y}}_{1}^{(1)}$; and, then, it inverts the resulting $2MT_2$ linearly independent equations relating the sent $2MT_2$ ${\tf{{v}}}_{11}$- and ${\tf{{v}}}_{21}$-symbols. Thus, Receiver 1 successfully decodes the ${\tf{{v}}}_{11}$- and ${\tf{{v}}}_{21}$-symbols that are intended to it. Receiver 2 performs similar operations to successfully decode the ${\tf{{v}}}_{12}$- and ${\tf{{v}}}_{22}$-symbols that are intended to it.
\vspace{.5em}

\noindent\textit{\textbf{Security Analysis}}

\noindent The analysis and algebra in this section are similar to in~\cite{YKPS11} in the context of secure broadcasting of messages on a two-user MIMO broadcast channel with delayed CSIT. 

\noindent At the end of Phase 4, the channel outputs at the receivers can be written as
\begin{align}
& {\tf{{y}}}_1 = \nonumber\\
& \underbrace{\left [
	\begin{matrix}
	\tilde{\tf{H}}_{2}  & \tilde{\tf{H}}_{11}^{(2)}\Theta_1 & \bf{0} \\
	\tilde{\tf{H}}_{4}\Phi_1\tilde{\tf{G}}_{2}  &  \tilde{\tf{H}}_{4}\Phi_1\tilde{\tf{H}}_{21}^{(2)}{\Theta}_1 & \tilde{\tf{H}}_{4}\Phi_2 \\
	\bf{0}  & {\bf{I}}_{NT_1}& \bf{0} \\
	\bf{0}  & \bf{0}  & {\bf{I}}_{NT_2} \\
 	\end{matrix} \right ]}_{\hat{\tf{H}} \:\in\: \mb{C}^{4M^2N\times 4M^2N}} \left [
 	\begin{matrix}
	{{\tf{v}}_1}   \\
	\tilde{\tf{H}}_{1} {{\tf{u}}}   \\
	\tilde{{\tf{H}}}_{3} {\tf{v}}_{2} + \tilde{{\tf{H}}}_{12}^{(3)}\Theta_2 \tilde{{{\tf{G}}}}_{1} {{\tf{u}}}
 	\end{matrix} \right ] \\
\label{main_equation2}
& {\tf{{y}}}_2 = \nonumber\\
& \underbrace{\left[
 	\begin{matrix}
	\bf{0}  & {\bf{I}}_{NT_1}& \bf{0} \\
	\bf{0}  & \bf{0}  & {\bf{I}}_{NT_2} \\
	\tilde{{\tf{G}}}_{3}  & \tilde{{\tf{H}}}_{22}^{(3)}\Theta_2 & \bf{0} \\
	\tilde{\tf{G}}_{4}\Phi_2\tilde{{\tf{H}}}_{3}  &  \tilde{\tf{G}}_{4}\Phi_2 \tilde{{\tf{H}}}_{12}^{(3)}\Theta_2 & \tilde{\tf{G}}_{4}\Phi_1  
	\end{matrix} \right ]}_{\hat{{\tf{G}}} \:\in\: \mb{C}^{4M^2N\times 4M^2N} } \left [
 	\begin{matrix}
	{\tf{v}}_{2}   \\
	\tilde{{\tf{G}}}_{1} {{\tf{u}}}   \\
	\tilde{{{\tf{G}}}}_{2} {{\tf{v}}_1}+\tilde{{\tf{H}}}_{21}^{(2)}{\Theta}_1\tilde{{{\tf{H}}}}_{1} {{\tf{u}}}
	\end{matrix} \right ]
\end{align}
where $\tilde{{{\tf{H}}}}_{t} = [\tilde{{{\tf{H}}}}_{11}^{(t)} \:\:\:\: \tilde{{{\tf{H}}}}_{12}^{(t)}], \:\: \tilde{{{\tf{G}}}}_{t} = [\tilde{{{\tf{H}}}}_{21}^{(t)} \:\:\:\: \tilde{{{\tf{H}}}}_{22}^{(t)}],\: \text{for}\: t = 1,\hdots,4$, ${{\tf{u}}} =[{{\tf{u}}}_{1}^T\:\: {{\tf{u}}}_{2}^T]^T,\:\: {\tf{v}}_{1} =[{{\tf{v}}}_{11}^T\:\: {{\tf{v}}}_{21}^T]^T,\:\text{and}\: {\tf{v}}_{2} =[{\tf{v}}_{12}^T\:\: {\tf{v}}_{22}^T]^T$. The information rate to Receiver 1 is given by the mutual information $I({\tf{v}}_1;{\tf{{y}}}_1)$, and can be evaluated as 
\setlength{\arraycolsep}{0.2em}
 \begin{align}
& I({{\tf{v}}}_1;{\tf{{y}}}_1)  = I({{\tf{v}}}_1,\tilde{{{\tf{H}}}}_{1} {{\tf{u}}},\tilde{{\tf{H}}}_{3} {\tf{v}}_{2} + \tilde{\tf{H}}_{12}^{(3)}\Theta_2 \tilde{\tf{G}}_{1} {\tf{u}};{\tf{{y}}}_1) \notag\\
& - I(\tilde{{{\tf{H}}}}_{1} {{\tf{u}}},\tilde{{\tf{H}}}_{3} {\tf{v}}_{2} + \tilde{{\tf{H}}}_{12}^{(3)}\Theta_2 \tilde{{{\tf{G}}}}_{1} {{\tf{u}}};{\tf{{y}}}_1|{\tf{v}}_{1}) \notag\\
& \overset{(a)}{=} \text{rank}(\hat{{\tf{H}}}). \log (2P) - \text{rank}\left (
 	\begin{matrix}
  	\tilde{{\tf{H}}}_{11}^{(2)}\Theta_1 & \bf{0} \\
  	\tilde{{\tf{H}}}_{4}\Phi_1\tilde{{\tf{H}}}_{21}^{(2)}{\Theta}_1 & \tilde{{\tf{H}}}_{4}\Phi_2 \\
  	{\bf{I}}_{N T_1}& \bf{0} \\
  	\bf{0}  & {\bf{I}}_{NT_2} \\
 	\end{matrix} \right ).\log(2P)\notag\\
&\overset{(b)}{=} N(T_1+T_2).\log(2P)+\text{rank}\left (
 	\begin{matrix}
	\tilde{{\tf{H}}}_{2}   \\
	\tilde{{\tf{H}}}_{4}\Phi_1\tilde{{{\tf{G}}}}_{2} \\
 	\end{matrix} \right ).\log(2P)\notag\\
&\:\:\:- N(T_1+T_2).\log(2P)\notag\\
&= \text{rank}\left (
 	\begin{matrix}
	\tilde{{\tf{H}}}_{2}   \\
	\tilde{{\tf{H}}}_{4}\Phi_1\tilde{{{\tf{G}}}}_{2} \\
 	\end{matrix} \right ).\log(2P)\notag\\
&\overset{(c)}{=} 2MN(2M-N).\log (2P)
\end{align}
where $(a)$ follows from \cite[Lemma 2]{YKPS11}; $(b)$ follows from the block diagonalization structure of $\hat{{\tf{H}}}$; and $(c)$ follows by reasoning as in  \cite{YKPS11} for the selection of $\Phi_1$ with appropriate rank such that the equality holds.

\noindent Similarly, the information leaked to Receiver 2 can be bounded as
\begin{align}
& I({\tf{v}}_1;{\tf{{y}}}_2)  = I({\tf{v}}_1;{\tf{{y}}}_2|{\tf{v}}_{2}) \leq I(\tilde{\tf{G}}_{2} {\tf{v}}_1;{\tf{y}}_2|{\tf{v}}_{2})\notag\\
&=  I(\tilde{{\tf{G}}}_{2} {{\tf{v}}}_1, {{\tf{u}}};{\tf{{y}}}_2|{\tf{v}}_{2}) - I({{\tf{u}}};{\tf{{y}}}_2|\tilde{{\tf{G}}}_{2} {{\tf{v}}}_1,{\tf{v}}_{2}) \notag\\
&\leq I(\tilde{{\tf{G}}}_{1} {{\tf{u}}} ,\tilde{{{\tf{G}}}}_{2} {{\tf{v}}}_1+\tilde{{\tf{H}}}_{21}^{(2)}{\Theta}_1\tilde{{{\tf{H}}}}_{1} {{\tf{u}}};{\tf{{y}}}_2|{\tf{v}}_{2})- I({{\tf{u}}};{\tf{{y}}}_2|\tilde{{\tf{G}}}_{2} {{\tf{v}}}_1,{\tf{v}}_{2}) \notag\\
&\overset{(a)}{=} \text{rank} \left (
 	\begin{matrix}
 	{\bf{I}}_{NT_1}& \bf{0} \\
 	\bf{0} & {\bf{I}}_{NT_2}  \\
 	\tilde{{\tf{H}}}_{22}^{(3)}\Theta_2 & \bf{0} \\  \tilde{{\tf{G}}}_{4}\Phi_2 \tilde{{\tf{H}}}_{12}^{(3)}\Theta_2 & \:\: \tilde{{\tf{G}}}_{4}\Phi_1  
 	\end{matrix} \right ).\log(2P) \notag\\
 & -\text{rank} \left (
 	\begin{matrix}
	\tilde{{\tf{G}}}_{1} \\
	\tilde{{\tf{H}}}_{21}^{(2)}{\Theta}_1\tilde{{{\tf{H}}}}_{1} \\
	\tilde{{\tf{H}}}_{22}^{(3)}\Theta_2\tilde{{\tf{G}}}_{1}  \\
 	\tilde{{\tf{G}}}_{4}\Phi_2 \tilde{{\tf{H}}}_{12}^{(3)}\Theta_2\tilde{{\tf{G}}}_{1}+ \tilde{{\tf{G}}}_{4}\Phi_1\tilde{{\tf{H}}}_{21}^{(2)}{\Theta}_1\tilde{{{\tf{H}}}}_{1}   
 \end{matrix} \right ). \log(2P)   \notag \\
&=  N(T_1+T_2). \log(2P) - \text{rank} \left (
 	\begin{matrix}
	\tilde{{\tf{G}}}_{1} \\
	\tilde{{\tf{H}}}_{21}^{(2)}{\Theta}_1\tilde{{{\tf{H}}}}_{1}
 	\end{matrix} \right ). \log(2P) \notag \\
&\overset{(b)}{=} 0
\end{align}
where $(a)$ follows from~\cite[Lemma 2]{YKPS11}; and $(b)$ follows by choosing $\Theta_1$ by reasoning similar to in~\cite{YKPS11}. 

\noindent From the above analysis, it can be easily seen that $2MN(2M-N)$ symbols are transmitted securely to Receiver 1 over a total of $4M^2$ time slots, thus yielding $d_{11}+d_{21}= N(2M-N)/{2M}$ sum SDoF at this receiver. Similar reasoning and algebra shows that $2MN(2M-N)$ symbols are also transmitted securely to Receiver 2 over a total of $4M^2$ time slots, thus yielding $d_{12}+d_{22}=N(2M-N)/{2M}$ sum SDoF at this receiver.
\setlength{\arraycolsep}{0.5em}

\subsection{Case 2: $2M \geq 2N$}\label{secIV_subsecB}

In this case, one can use the coding scheme of Section~\ref{secIV_subsecA}, with each transmitter utilizing only $N$ antennas among the $M$ antennas with which it is equipped. In what follows, we briefly describe an alternate coding scheme in which Receiver $i$, $i=1,2$, feeds back only its output to transmitter $i$, i.e., delayed CSI is not required. Also, as it will be seen from what follows, this coding scheme requires a shorter time delay comparatively. Some of the details of the analysis of this coding scheme are similar to in  Section~\ref{secIV_subsecA}, however; and so we only outline them briefly. More specifically, the communication takes place in four phases, each composed of only one time slot.
\vspace{.5em}

\noindent\textit{\textbf{Phase 1: Injecting artificial noise}}

\noindent In this phase, both transmitters inject artificial noise. Let ${\tf{{u}}}_{1} = [u^1_{1},\hdots,u^{N}_{1}]^T$ denote the artificial noise injected by Transmitter 1, and ${\tf{{u}}}_{2} = [u^1_{2},\hdots,u^{N}_{2}]^T$ denote the  artificial noise injected by Transmitter 2. The channel outputs at the receivers during this phase are given by
\begin{eqnarray}
{\bf{y}}_{1}^{(1)} = {\bf{H}}_{11}^{(1)} {\tf{u}}_1+  {\bf{H}}_{12}^{(1)} {\tf{u}}_2 \\
{\bf{y}}_{2}^{(1)} = {\bf{H}}_{21}^{(1)} {\tf{u}}_1+  {\bf{H}}_{22}^{(1)} {\tf{u}}_2
\end{eqnarray}
where ${\bf{H}}_{ji}^{(1)} \in \mb{C}^{N \times N}$, for $i = 1,2$, $j = 1,2$, ${\bf{y}}_{1}^{(1)} \in \mb{C}^{N}$ and ${\bf{y}}_{2}^{(1)} \in \mb{C}^{N}$. At the end of this phase, the output at Receiver $i$, $i=1,2$, is fed back to Transmitter $i$. 
\vspace{.5em}

\noindent\textit{\textbf{Phase 2: Fresh information for Receiver 1}}

\noindent In this phase, both transmitters transmit confidential messages to Receiver 1. These messages are meant to be concealed from Receiver 2. To this end, Transmitter 1 transmits fresh information ${\tf{{v}}}_{11} = [v^1_{11},\hdots,v^{N}_{11}]^T$ along with a linear combination of the channel output at Receiver 1 during Phase 1, and Transmitter 2 transmits fresh information ${\tf{{v}}}_{21} = [v^1_{21},\hdots,v^{N}_{21}]^T$ intended for Receiver 1, i.e.,
\begin{eqnarray}
&&{\tf{x}}_1 = {\tf{v}}_{11}+ \Theta_1{\bf{y}}_{1}^{(1)} \notag\\
&&{\tf{x}}_2 = {\tf{v}}_{21}
\end{eqnarray}
where $\Theta_1 \in \mc{C}^{N \times N}$ is a matrix that is assumed to be known at all the nodes, and whose choice will be specified below. The channel outputs at the receivers during this phase are given by
\begin{subequations}
\begin{align}
{\bf{y}}_{1}^{(2)} &=  {\bf{H}}_{11}^{(2)} ({\tf{v}}_{11}+ \Theta_1 {\bf{y}}_{1}^{(1)})+  {\bf{H}}_{12}^{(2)} {\tf{v}}_{21} \\
{\bf{y}}_{2}^{(2)} &=  {\bf{H}}_{21}^{(2)} ({\tf{v}}_{11}+ \Theta_1 {\bf{y}}_{1}^{(1)})+  {\bf{H}}_{22}^{(2)} {\tf{v}}_{21} 
\end{align}
\end{subequations}
\noindent where ${\bf{H}}_{ji}^{(2)} \in \mb{C}^{N \times N}$, for $i = 1,2$, $j = 1,2$, ${\bf{y}}_{1}^{(2)} \in \mb{C}^{N}$ and ${\bf{y}}_{2}^{(2)} \in \mb{C}^{N}$. At the end of this phase, the channel output at Receiver $i$, $i=1,2$, is fed back to Transmitter $i$. Since Receiver 1 knows the CSI and the channel output ${\tf{{y}}}^{(1)}_1$ from Phase 1, it subtracts out the contribution of ${\tf{{y}}}^{(1)}_1$ from ${\bf{y}}_{1}^{(2)}$ and, thus, obtains $N$ linearly independent equations that relates the $2N$ ${\tf{{v}}}_{11}$- and ${\tf{{v}}}_{21}$-symbols. Thus, Receiver 1 requires $N$ extra linearly independent equations to successfully decode the ${\tf{{v}}}_{11}$- and ${\tf{{v}}}_{21}$-symbols that are intended to it during this phase. These extra equations will be provided by transmitting ${\bf{y}}_{2}^{(2)}$ by Transmitter 2 in Phase 4. Transmitter 2 learns ${\tf{{y}}}^{(2)}_2$ directly by means of the output feedback from Receiver 2 at the end of this phase. 
\vspace{.5em}

\noindent\textit{\textbf{Phase 3: Fresh information for Receiver 2}}

\noindent This phase is similar to Phase 2, with the roles of Transmitter 1 and Transmitter 2, as well as those of Receiver 1 and Receiver 2, being swapped. The information messages are sent by both transmitters to Receiver 2, and are to be concealed from Receiver 1. More specifically, Transmitter 1 transmits fresh information ${\tf{{v}}}_{12} = [v^1_{12},\hdots,v^{N}_{12}]^T$ to Receiver 2, and Transmitter 2 transmits ${\tf{{v}}}_{22} = [v^1_{22},\hdots,v^{N}_{22}]^T$ along with a linear combination of the channel output received at Receiver 2 during Phase 1, i.e.,
\begin{eqnarray}
&&{\tf{x}}_1= {\tf{v}}_{12} \notag\\
&&{\tf{x}}_2 = {\tf{v}}_{22} + \Theta_2 {\bf{y}}_{2}^{(1)}
\end{eqnarray}
where $\Theta_2 \in \mc{C}^{N \times N }$ is matrix that is known at all nodes and whose choice will be specified below. The channel outputs at the receivers during this phase are given by
\begin{subequations}
\begin{align}
{\bf{y}}_{1}^{(3)} &= {\bf{H}}_{11}^{(3)} {\tf{v}}_{21}+ {\bf{H}}_{12}^{(3)} ({\tf{v}}_{22}+ \Theta_2 {\bf{y}}_{2}^{(1)}) \\
{\bf{y}}_{2}^{(3)} &= {\bf{H}}_{21}^{(3)} {\tf{v}}_{21}+ {\bf{H}}_{22}^{(3)} ({\tf{v}}_{22}+ \Theta_2 {\bf{y}}_{2}^{(1)})
\end{align}
\end{subequations}
\noindent where ${\bf{H}}_{ji}^{(3)} \in \mb{C}^{N \times N}$, for $i = 1, 2$, $j = 1, 2$, ${\bf{y}}_{1}^{(3)} \in \mb{C}^{N}$ and ${\bf{y}}_{2}^{(3)} \in \mb{C}^{N}$. At the end of this phase, the channel output at Receiver $i$, $i=1,2$, is fed back to Transmitter $i$. Since Receiver 2 knows the CSI and the channel output ${\tf{{y}}}^{(1)}_2$ from Phase 1, it subtracts out the contribution of ${\tf{{y}}}^{(1)}_2$ from ${\bf{y}}_{2}^{(3)}$ and, thus, obtains $N$ linearly independent equations that relates the $2N$ ${\tf{{v}}}_{21}$- and ${\tf{{v}}}_{22}$-symbols. Thus, Receiver 2 requires $N$ extra linearly independent equations to successfully decode the ${\tf{{v}}}_{21}$- and ${\tf{{v}}}_{22}$-symbols that are intended to it during this phase. These extra equations will be provided by transmitting ${\bf{y}}_{1}^{(3)}$ by Transmitter 1 in Phase 4. Transmitter 1 learns ${\tf{{y}}}^{(3)}_1$ directly by means of the output feedback from Receiver 1 at the end of this phase. 
\vspace{.5em}

\noindent\textit{\textbf{Phase 4: Interference alignment and decoding}}

\noindent Recall that, at the end of Phase 3, Receiver 1 knows ${\tf{{y}}}^{(3)}_1$ and requires ${\tf{{y}}}^{(2)}_2$; and Receiver 2  knows ${\tf{{y}}}^{(2)}_2$ and requires ${\tf{{y}}}^{(3)}_1$. Also, at the end of this phase, Transmitter 1 has learned ${\tf{{y}}}^{(3)}_1$ by means of output feedback from Receiver 1; and Transmitter 2  has learned ${\tf{{y}}}^{(2)}_2$ by means of output feedback from Receiver 2. The inputs by the two transmitters during Phase 4 are given by 
\begin{eqnarray}
&&{\tf{{x}}}_1 = \Phi_2 {\tf{{y}}}^{(3)}_1 \notag\\
&&{\tf{{x}}}_2 = \Phi_1 {\tf{{y}}}^{(2)}_2
\end{eqnarray} 
\noindent where $\Phi_{1} \in \mc{C}^{N \times N}$ and $\Phi_{2} \in  \mc{C}^{N \times N}$ are matrices that are assumed to be known by all the nodes. At the end of Phase 4, Receiver 1 gets $N$ equations in $2N$ variables. Since Receiver 1 knows ${\tf{{y}}}^{(3)}_1$, as well as the CSI, it can subtract out the side information, or interference, equations ${\tf{{y}}}^{(2)}_2$ that are seen at Receiver 2 during Phase 2. Then, using the pair of output vectors $({\bf{y}}_{1}^{(2)}, {\bf{y}}_{2}^{(2)})$, Receiver 1 first subtracts out the contribution of ${\bf{y}}_{1}^{(1)}$; and, then, it inverts the resulting $2N$ linearly independent equations relating the sent $2N$ ${\tf{{v}}}_{11}$- and ${\tf{{v}}}_{21}$-symbols. Thus, Receiver 1 successfully decodes the ${\tf{{v}}}_{11}$- and ${\tf{{v}}}_{21}$-symbols that are intended to it. Receiver 2 performs similar operations to successfully decode the ${\tf{{v}}}_{12}$- and ${\tf{{v}}}_{22}$-symbols that are intended to it.

\noindent\textit{\textbf{Security Analysis}}

\noindent At the end of Phase 4, the channel outputs at the receivers are given by
\begin{align}
& {\tf{{y}}}_1 = \nonumber\\
& \underbrace{\left [
 	\begin{matrix}
	{\tf{H}}_{2}  & {\tf{H}}_{11}^{(2)}\Theta_1 & \bf{0} \\
	{\tf{H}}_{12}^{(4)}\Phi_1{\tf{G}}_{2}  &  {\tf{H}}_{12}^{(4)}\Phi_1{\tf{H}}_{21}^{(2)}{\Theta}_1 & {\tf{H}}_{11}^{(4)}\Phi_2 \\
	\bf{0}  & {\bf{I}}_{N}& \bf{0} \\
	\bf{0}  & \bf{0}  & {\bf{I}}_{N} \\
 	\end{matrix} \right ]}_{\hat{\tf{H}} \:\in\: \mb{C}^{4N\times 4N} } \left [
 	\begin{matrix}
	{{\tf{v}}_1}   \\
	{\tf{H}}_{1} {{\tf{u}}}   \\
	{{\tf{H}}}_{3} {\tf{v}}_{2} + {{\tf{H}}}_{12}^{(3)}\Theta_2 {{{\tf{G}}}}_{1} {{\tf{u}}}
 	\end{matrix} \right ] \\
& {\tf{{y}}}_2 = \nonumber\\
& \underbrace{\left[
 	\begin{matrix}
	\bf{0}  & {\bf{I}}_{N}& \bf{0} \\
	\bf{0}  & \bf{0}  & {\bf{I}}_{N} \\
	{{\tf{G}}}_{3}  & {{\tf{H}}}_{22}^{(3)}\Theta_2 & \bf{0} \\
	{\tf{H}}_{21}^{(4)}\Phi_2{{\tf{H}}}_{3}  &  {\tf{H}}_{21}^{(4)}\Phi_2 {{\tf{H}}}_{12}^{(3)}\Theta_2 & {\tf{H}}_{22}^{(4)}\Phi_1  
 	\end{matrix} \right ]}_{\hat{{\tf{G}}} \:\in\: \mb{C}^{4N \times 4N } } \left [
 	\begin{matrix}
	{\tf{v}}_{2}   \\
	{{\tf{G}}}_{1} {{\tf{u}}}   \\
	{{{\tf{G}}}}_{2} {{\tf{v}}_1}+{{\tf{H}}}_{21}^{(2)}{\Theta}_1{{{\tf{H}}}}_{1} {{\tf{u}}}
 	\end{matrix} \right ]
\end{align}
where ${{{\tf{H}}}}_{t} = [{{{\tf{H}}}}_{11}^{(t)} \:\:\:\: {{{\tf{H}}}}_{12}^{(t)}], \:\: {{{\tf{G}}}}_{t} = [{{{\tf{H}}}}_{21}^{(t)} \:\:\:\: {{{\tf{H}}}}_{22}^{(t)}],\: \text{for}\: t =1,\hdots,3$, ${{\tf{u}}} =[{{\tf{u}}}_{1}^T\:\: {{\tf{u}}}_{2}^T]^T,\:\: {\tf{v}}_{1} =[{{\tf{v}}}_{11}^T\:\: {{\tf{v}}}_{21}^T]^T,\: \text{and} \: {\tf{v}}_{2} =[{\tf{v}}_{12}^T\:\: {\tf{v}}_{22}^T]^T$. Similar to the analysis of the previous case, the information rate to Receiver 1 is given by the mutual information $I({\tf{v}}_1;{\tf{{y}}}_1)$, and can be evaluated as 
\setlength{\arraycolsep}{.2em}
\begin{align}
& I({{\tf{v}}}_1;{\tf{{y}}}_1) = I({{\tf{v}}}_1,{{{\tf{H}}}}_{1} {{\tf{u}}},{{\tf{H}}}_{3} {\tf{v}}_{2} + {\tf{H}}_{12}^{(3)}\Theta_2 {\tf{G}}_{1} {\tf{u}};{\tf{{y}}}_1) \notag\\
& - I({{{\tf{H}}}}_{1} {{\tf{u}}},{{\tf{H}}}_{3} {\tf{v}}_{2} + {{\tf{H}}}_{12}^{(3)}\Theta_2 {{{\tf{G}}}}_{1} {{\tf{u}}};{\tf{{y}}}_1|{\tf{v}}_{1}) \notag\\
& \overset{(a)}{=} \text{rank}(\hat{{\tf{H}}}). \log (2P) \notag\\
& - \text{rank}\left (
 	\begin{matrix}
  	{{\tf{H}}}_{11}^{(2)}\Theta_1 & \bf{0} \\
  	{{\tf{H}}}_{12}^{(4)}\Phi_1{{\tf{H}}}_{21}^{(2)}{\Theta}_1 & {{\tf{H}}}_{11}^{(4)}\Phi_2 \\
  	{\bf{I}}_{N }& \bf{0} \\
  	\bf{0}  & {\bf{I}}_{N} \\
 	\end{matrix} \right ).\log(2P)\notag\\
&\overset{(b)}{=} 2N.\log(2P)+\text{rank}\left (
 	\begin{matrix}
	{{\tf{H}}}_{2}   \\
	{{\tf{H}}}_{12}^{(4)}\Phi_1{{{\tf{G}}}}_{2} \\
 	\end{matrix} \right ).\log(2P)- 2N.\log(2P)\notag\\
&= \text{rank}\left (
 	\begin{matrix}
	{{\tf{H}}}_{2}   \\
	{{\tf{H}}}_{12}^{(4)}\Phi_1{{{\tf{G}}}}_{2} \\
 	\end{matrix} \right ).\log(2P)\notag\\
&\overset{(c)}{=} 2N.\log (2P)
\end{align}
where $(a)$ follows from \cite[Lemma 2]{YKPS11}; $(b)$ follows by using the block diagonalization structure of $\hat{{\tf{H}}}$; and $(c)$ follows by  reasoning as in \cite{YKPS11} for the selection of $\Phi_1$ with appropriate rank such that the equality holds.

\noindent Similarly, the information leaked to Receiver 2  can be bounded as
\begin{align}
& I({\tf{v}}_1;{\tf{{y}}}_2)  \notag\\
& \leq I({{\tf{G}}}_{1} {{\tf{u}}} ,{{{\tf{G}}}}_{2} {{\tf{v}}}_1+{{\tf{H}}}_{21}^{(2)}{\Theta}_1{{{\tf{H}}}}_{1} {{\tf{u}}};{\tf{{y}}}_2|{\tf{v}}_{2})- I({{\tf{u}}};{\tf{{y}}}_2|{{\tf{G}}}_{2} {{\tf{v}}}_1,{\tf{v}}_{2}) \notag\\
& \overset{(a)}{=} \text{rank} \left (
 	\begin{matrix}
 	{\bf{I}}_{N}& \bf{0} \\
 	\bf{0} & {\bf{I}}_{N}  \\
 	{{\tf{H}}}_{22}^{(3)}\Theta_2 & \bf{0} \\  {{\tf{H}}}_{21}^{(4)}\Phi_2 {{\tf{H}}}_{12}^{(3)}\Theta_2 & \:\: {{\tf{H}}}_{22}^{(4)}\Phi_1  
 	\end{matrix} \right ).\log(2P) \notag\\
& -\text{rank} \left (
 	\begin{matrix}
	{{\tf{G}}}_{1} \\
	{{\tf{H}}}_{21}^{(2)}{\Theta}_1{{{\tf{H}}}}_{1} \\
	{{\tf{H}}}_{22}^{(3)}\Theta_2{{\tf{G}}}_{1}  \\
	{{\tf{H}}}_{21}^{(4)}\Phi_2 {{\tf{H}}}_{12}^{(3)}\Theta_2{{\tf{G}}}_{1}+ {{\tf{H}}}_{22}^{(4)}\Phi_1{{\tf{H}}}_{21}^{(2)}{\Theta}_1{{{\tf{H}}}}_{1}   
 	\end{matrix} \right ). \log(2P)   \notag \\
&= 2N. \log(2P) - \text{rank} \left (
 	\begin{matrix}
	{{\tf{G}}}_{1} \\
	{{\tf{H}}}_{21}^{(2)}{\Theta}_1{{{\tf{H}}}}_{1}
 	\end{matrix} \right ). \log(2P) \notag \\
&\overset{(b)}{=} 0
\end{align}
where $(a)$ follows from \cite[Lemma 2]{YKPS11};  and $(b)$ follows by choosing $\Theta_1$ with the reasoning similar to \cite{YKPS11}. 

\noindent From the above analysis, it can be easily seen that $2N$ symbols are transmitted securely to Receiver 1, over a total of $4$ time slots, yielding $d_{11}+d_{21}= N/2$ sum SDoF. Similar analysis shows that the scheme also offers $d_{12}+d_{22}= N/2$ sum SDoF for Receiver 2. 

\noindent This concludes the proof of the direct part of Theorem~\ref{theorem-sum-sdof-region-mimo-x-channel-with-asymmetric-output-feedback-and-delayed-csi}.

\vspace{.5em}
\begin{remark}\label{role-of-feedback-coding-scheme-of-theorem1}
Investigating the coding scheme of Theorem~\ref{theorem-sum-sdof-region-mimo-x-channel-with-asymmetric-output-feedback-and-delayed-csi}, it can be seen that in the case in which $N \leq M$, asymmetric output feedback only suffices to achieve the optimum sum SDoF point. That is, the transmitters exploit only the availability of asymmetric output feedback, and do not make use of the available delayed CSIT.
\end{remark}

\section{SDoF of MIMO X-channel with Only Output Feedback}\label{secV}

In this section, we focus on the two-user MIMO X-channel with only feedback available at transmitters. We study two special cases of availability of feedback at transmitters, 1) the case in which each receiver feeds back its channel output to both transmitters, i.e., \textit{symmetric output feedback}, and 2) the case in which Receiver $i$, $i=1,2$, feeds back its output only to Transmitter $i$, i.e., \textit{asymmetric output feedback}. In both cases, no CSI is provided to the transmitters. The model with symmetric output feedback may model a setting in which both feedback signals are strong and can be heard by both transmitters. The model with asymmetric output feedback may model a setting in which the feedback signals are weak and can be heard by only one transmitter each.

\subsection{MIMO X-channel with symmetric output feedback}\label{secV_subsecA}

\vspace{.5em}
\noindent The following theorem provides the sum SDoF region of the MIMO X-channel with symmetric output feedback.
\vspace{.5em}
\begin{theorem}\label{theorem-sum-sdof-region-mimo-x-channel-with-symmetric-output-feedback}
The sum SDoF region of the two-user $(M,M,N,N)$--MIMO X-channel with symmetric output feedback is given by that of Theorem~\ref{theorem-sum-sdof-region-mimo-x-channel-with-asymmetric-output-feedback-and-delayed-csi}.
\end{theorem}

\vspace{.5em}
\begin{remark}\label{remark-symmetric-feedback}
The sum SDoF region of the MIMO X-channel with symmetric output feedback is same as the sum SDoF region of the MIMO X-channel with asymmetric output feedback and delayed CSIT. Investigating the coding scheme of the MIMO X-channel with asymmetric output feedback and delayed CSIT of Theorem~\ref{theorem-sum-sdof-region-mimo-x-channel-with-asymmetric-output-feedback-and-delayed-csi}, it can be seen that the delayed CSIT is utilized therein to provide each transmitter with the equations (or, side information) that are heard at the other receiver, which is unintended. With the availability of the output feedback symmetrically, this information is readily available at each transmitter; and, thus, there is no need for any CSIT at the transmitters in order to achieve the same sum SDoF region as that of  Theorem~\ref{theorem-sum-sdof-region-mimo-x-channel-with-asymmetric-output-feedback-and-delayed-csi}.
\end{remark}
\vspace{.5em}
\begin{IEEEproof}
 The proof of the outer bound can be obtained by reasoning as follows. Let us denote the two-user MIMO X-channel with symmetric output feedback that we study as $\text{MIMO-X}^{(0)}$. Consider the MIMO X-channel obtained by assuming that, in addition to symmetric output feedback, i) delayed CSIT is provided to both transmitters and that ii) the transmitters are allowed to cooperate. Denote the obtained MIMO X-channel as  $\text{MIMO-X}^{(1)}$. Since the transmitters cooperate in $\text{MIMO-X}^{(1)}$, this model is in fact a MIMO BC with $2M$ antennas at the transmitter and $N$ antennas at each receiver, with delayed CSIT as well as output feedback given to the transmitter. Then, an outer bound on the SDoF of this $\text{MIMO-X}^{(1)}$ is given by ~\cite[Theorem 3]{YKPS11}. This holds because the result of~\cite[Theorem 3]{YKPS11} continues to hold if one provides outputs feedback from the receivers to the transmitter in the two-user MIMO BC with delayed CSIT that is considered in~\cite{YKPS11}.  Next, since delayed CSIT at the transmitters and cooperation can only increase the SDoF, it follows that the obtained outer bound is also an outer bound on the SDoF of $\text{MIMO-X}^{(0)}$. Thus, the region of Theorem~\ref{theorem-sum-sdof-region-mimo-x-channel-with-asymmetric-output-feedback-and-delayed-csi} is an outer bound on the sum SDoF region for the MIMO X-channel in which the transmitters are provided only with symmetric output feedback. 
 
We now provide a brief outline of the coding scheme that we use to establish the sum SDoF region of Theorem~\ref{theorem-sum-sdof-region-mimo-x-channel-with-symmetric-output-feedback}. This coding scheme is very similar to that we use for the proof of Theorem~\ref{theorem-sum-sdof-region-mimo-x-channel-with-asymmetric-output-feedback-and-delayed-csi}, with the following (rather minor) differences. For the case in which $2M \leq N$ and that in which $2N \leq 2M$, the coding strategies are  exactly same as those that we used for the proof of Theorem~\ref{theorem-sum-sdof-region-mimo-x-channel-with-asymmetric-output-feedback-and-delayed-csi}. For the case in which $N \leq 2M \leq 2N$, the first three phases are similar to those in the coding scheme of Theorem~\ref{theorem-sum-sdof-region-mimo-x-channel-with-asymmetric-output-feedback-and-delayed-csi}, but with, at the end of these phases, the receivers feeding back their outputs to both transmitters, instead of Receiver $i$, $i=1,2$, feeding back its output together with the delayed CSI to Transmitter $i$. Note that, during these phases, each transmitter learns the required side information equations \textit{directly} from the symmetric output feedback that it gets from the receivers (see Remark~\ref{remark-symmetric-feedback}). Phase 4 and the decoding procedures are similar to those in the proof of Theorem~\ref{theorem-sum-sdof-region-mimo-x-channel-with-asymmetric-output-feedback-and-delayed-csi}. This concludes the proof of Theorem~\ref{theorem-sum-sdof-region-mimo-x-channel-with-symmetric-output-feedback}.
\end{IEEEproof}

\subsection{MIMO X-channel with only asymmetric output feedback}\label{secV_subsecB}

We now consider the case in which only asymmetric output feedback is provided from the receivers to the transmitters, i.e., Receiver $i$, $i=1,2$, feeds back its output to only Transmitter $i$. 

\noindent For convenience, we define the following quantity. Let, for given non-negative $(M,N)$,
\begin{equation}
d^{\text{local}}_s(N,N,M)=\left\{
\begin{array}{ll}
0 & \:\:\: \text{if} \:\:\: M \leq N\\
\frac{M^2(M-N)}{2N^2+(M-N)(3M-N)} & \:\:\:\text{if} \:\:\: N \leq M \leq 2N\\
\frac{2N}{3} &\:\:\: \text{if} \:\:\: M \geq 2N
\end{array}
\right.
\end{equation}

\noindent The following theorem provides an inner bound on the sum SDoF region of the two-user MIMO X-channel with asymmetric output feedback.
\vspace{.5em}
\begin{theorem}\label{theorem-sum-sdof-region-mimo-x-channel-with-asymmetric-output-feedback}
An inner bound on the sum SDoF region of the two-user $(M,M,N,N)$--MIMO X-channel with asymmetric output feedback is given by the set of all non-negative pairs $(d_{11}+d_{21},d_{12}+d_{22})$ satisfying
\begin{align}
\frac{d_{11}+d_{21}}{d^{\text{local}}_s(N,N,2M)}+\frac{d_{12}+d_{22}}{\min(2M,2N)} &\leq 1 \nonumber\\
\frac{d_{11}+d_{21}}{\min(2M,2N)}+\frac{d_{12}+d_{22}}{d^{\text{local}}_s(N,N,2M)} &\leq 1
\label{linear-equations-corner-points-theorem-sum-sdof-region-mimo-x-channel-with-asymmetric-output-feedback}
\end{align}
for $2M \geq N$; and $\mc C^{\text{sum}}_{\text{SDoF}}=\{(0,0)\}$ if $2M \leq N$.
\end{theorem}
\vspace{.5em}
\begin{remark}
Obviously, the region of Theorem~\ref{theorem-sum-sdof-region-mimo-x-channel-with-asymmetric-output-feedback-and-delayed-csi} is an outer bound on the sum SDoF region of the MIMO X-channel with asymmetric output feedback. Also, it is easy to see that the inner bound of Theorem~\ref{theorem-sum-sdof-region-mimo-x-channel-with-asymmetric-output-feedback} is tight in the case in which $M \geq N$. 
\end{remark}
\vspace{.5em}
\begin{remark}\label{suboptimality-of-asymmetric-feedback}
The main reason for which the inner bound of Theorem~\ref{theorem-sum-sdof-region-mimo-x-channel-with-asymmetric-output-feedback} is smaller than that of Theorem~\ref{theorem-sum-sdof-region-mimo-x-channel-with-asymmetric-output-feedback-and-delayed-csi} for the model with asymmetric output feedback and delayed CSIT can be explained as follows. Consider the Phase 4 in the coding scheme of Theorem~\ref{theorem-sum-sdof-region-mimo-x-channel-with-asymmetric-output-feedback-and-delayed-csi} in Section~\ref{secIV_subsecB}. Each receiver requires $N(2M-N)(2M-N)$ extra equations to decode the symbols that are intended to it correctly. Given that there are more equations that need to be transmitted to both receivers than the number of available antennas at the transmitters, some of the equations need to be sent by both transmitters, i.e., some of the available antennas send sums of two equations, one intended for each receiver. Then, it can be seen easily that this is only possible if both transmitters know the ensemble of side information equations that they need to transmit, i.e., not only a subset of them corresponding to one receiver. In the coding scheme of Theorem~\ref{theorem-sum-sdof-region-mimo-x-channel-with-asymmetric-output-feedback-and-delayed-csi}, this is made possible by means of availability of both asymmetric output feedback and delayed CSIT. Similarly, in the coding scheme of Theorem~\ref{theorem-sum-sdof-region-mimo-x-channel-with-symmetric-output-feedback}, this is made possible by means of availability of symmetric output feedback at the transmitters. For the model with only asymmetric output feedback, however, it is not clear how this can be obtained (if possible at all); and this explains the loss incurred in the sum SDoF region. More specifically, consider Phase 2 of the coding scheme of Theorem~\ref{theorem-sum-sdof-region-mimo-x-channel-with-asymmetric-output-feedback-and-delayed-csi}. Recall that, at the beginning of this phase, Transmitter 1 utilizes the fed back CSI $(\tilde{\bf{H}}_{11}^{(2)},\tilde{\bf{H}}_{12}^{(2)})$ to learn the ${\tf{{v}}}_{21}$-symbols that are transmitted by Transmitter 2 during this phase; and then utilizes the fed back CSI $(\tilde{\bf{H}}_{21}^{(2)},\tilde{\bf{H}}_{22}^{(2)})$ to reconstruct the side information output vector ${\tf{{y}}}^{(2)}_2$ that is required by Receiver 1 (given by \eqref{output-receiver2-phase2-proof-of-theorem1}). Also, Transmitter 2 performs similar operations to learn the side information output vector ${\tf{{y}}}^{(3)}_1$ that is required by Receiver 2 (given by \eqref{output-receiver1-phase3-proof-of-theorem1}). In the case of only asymmetric output feedback given to the transmitters, as we mentioned previously, it is not clear whether this could be possible because of the lack of availability of CSIT. 
\end{remark}
\vspace{.5em}
\begin{IEEEproof}
We now provide an outline of the coding scheme for the MIMO X-channel with asymmetric output feedback.

\noindent For the case in which $2M \leq N$ and the case in which $N \leq M$, the achievability follows trivially by using the coding scheme of Theorem~\ref{theorem-sum-sdof-region-mimo-x-channel-with-asymmetric-output-feedback-and-delayed-csi} (see Remark~\ref{role-of-feedback-coding-scheme-of-theorem1}).

\noindent For the case in which $N \leq 2M \leq 2N$, the proof of achievability follows by a variation of the coding scheme of Theorem~\ref{theorem-sum-sdof-region-mimo-x-channel-with-asymmetric-output-feedback-and-delayed-csi} that we outline briefly in what follows. The communication takes place in four phases. 
\vspace{.5em}

\noindent \textbf{\textit{Phase 1:}} The transmission scheme in this phase is similar to that in Phase 1 of the coding scheme of Theorem~\ref{theorem-sum-sdof-region-mimo-x-channel-with-asymmetric-output-feedback-and-delayed-csi}, but with at the end of this phase, Receiver $i$, $i=1,2$, feeding back only its output to Transmitter $i$, instead of feeding back its output together with the delayed CSI to Transmitter $i$.
\vspace{.5em}

\noindent \textbf{\textit{Phase 2:}} The communication takes place in $T_2=M(2M-N)$ channel uses. The transmission scheme is same as that of Phase 2 of the coding scheme of Theorem~\ref{theorem-sum-sdof-region-mimo-x-channel-with-asymmetric-output-feedback-and-delayed-csi}, with the following modifications. The inputs $({\tf{x}}_1,{\tf{x}}_2)$ from the transmitters and outputs $({\tf{y}}^{(2)}_1,{\tf{y}}^{(2)}_2)$ at the receivers are again given by \eqref{inputs-phase2-proof-of-theorem1} and \eqref{outputs-receivers-phase2-proof-of-theorem1}, respectively. At the end of these phases, Receiver $i$, $i=1,2$, feeds back its output to Transmitter $i$. At the end of this phase, Receiver 1 requires $(2M-N)T_2$ extra linearly independent equations to successfully decode the ${\tf{{v}}}_{11}$- and ${\tf{{v}}}_{21}$-symbols that are intended to it during this phase. Let $\tilde{\bf{y}}_{2}^{(2)} \in \mb{C}^{(2M-N)T_2}$ denote a set of $(2M-N)T_2$ such linearly independent equations, selected among the available $NT_2$ side information equations ${\bf{y}}_{2}^{(2)} \in \mb{C}^{NT_2}$ (recall that $2M-N \leq N$ in this case). If these equations can be conveyed to Receiver 1, they will suffice to help it decode the ${\tf{{v}}}_{11}$- and ${\tf{{v}}}_{21}$-symbols, since the latter already knows ${\tf{{y}}}^{(1)}_1$. These equations will be transmitted by (only) Transmitter 2 in Phase 4. Transmitter 2 learns ${\tf{{y}}}^{(2)}_2$, and so $\tilde{\bf{y}}_{2}^{(2)}$, directly by means of the output feedback from Receiver 2 at the end of this phase.
\vspace{.5em}

\noindent \textbf{\textit{Phase 3:}} The communication takes place in $T_2=M(2M-N)$ channel uses. The transmission scheme is same as that of Phase 3 of the coding scheme of Theorem~\ref{theorem-sum-sdof-region-mimo-x-channel-with-asymmetric-output-feedback-and-delayed-csi}, with the following modifications. The inputs $({\tf{x}}_1,{\tf{x}}_2)$ from the transmitters and outputs $({\tf{{y}}}^{(2)}_1,{\tf{y}}^{(2)}_2)$ at the receivers are again given by \eqref{inputs-phase3-proof-of-theorem1} and \eqref{outputs-receivers-phase3-proof-of-theorem1}, respectively. At the end of this phase, Receiver 2 requires $(2M-N)T_2$ extra linearly independent equations to successfully decode the ${\tf{{v}}}_{12}$- and ${\tf{{v}}}_{22}$-symbols that are intended to it during this phase. Let $\tilde{\bf{y}}_{1}^{(3)} \in \mb{C}^{(2M-N)T_2}$ denote a set of $(2M-N)T_2$ such linearly independent equations, selected among the available $NT_2$ side information equations ${\bf{y}}_{1}^{(3)} \in \mb{C}^{NT_2}$ (recall that $2M-N \leq N$ in this case). These equations will be transmitted by (only) Transmitter 1 in Phase 4. Transmitter 1 learns ${\tf{{y}}}^{(3)}_1$, and so $\tilde{\bf{y}}_{1}^{(3)}$, directly by means of the output feedback from Receiver 1 at the end of this phase.
\vspace{.5em}

\noindent \textbf{\textit{Phase 4:}} Recall that at the end of Phase 3, Receiver 1 requires the side information output vector $\tilde{\bf{y}}_{2}^{(2)}$, and Receiver 2 requires the side information output vector $\tilde{\bf{y}}_{1}^{(3)}$. In Phase 4, the communication takes place in $T_3=(2M-N)(2M-N)$ channel uses. During this phase, Transmitter 1 transmits ${\tf{{x}}}_1 = \Phi_2 {\tf{{y}}}^{(3)}_1$  and Transmitter 2 transmits ${\tf{{x}}}_2 = \Phi_1 {\tf{{y}}}^{(2)}_2$, where $\Phi_{1} \in \mc{C}^{MT_3 \times NT_2}$, and  $\Phi_{2} \in  \mc{C}^{MT_3 \times NT_2}$, in $T_3$ channel uses. 
\vspace{.5em}

\noindent \textbf{\textit{Decoding:}} At the end of Phase 4, Receiver 1 gets $NT_3$ equations in $2MT_3$ variables. Since Receiver 1 knows  ${\bf{y}}_{1}^{(3)}$ from Phase 3 as well as the CSI, it can subtract out the contribution of $\tilde{\bf{y}}_{1}^{(3)}$ from its received signal to obtain the side information output vector $\tilde{\bf{y}}_{2}^{(2)}$. Then, using the pair of output vectors $({\bf{y}}_{1}^{(2)},\tilde{\bf{y}}_{2}^{(2)})$, Receiver 1 first subtracts out the contribution of ${\bf{y}}_{1}^{(1)}$; and, then, it inverts the resulting $2MT_2$ linearly independent equations relating the sent $2MT_2$ ${\tf{{v}}}_{11}$- and ${\tf{{v}}}_{21}$-symbols. Thus, Receiver 1 successfully decodes the ${\tf{{v}}}_{11}$- and ${\tf{{v}}}_{21}$-symbols that are intended to it. Receiver 2 performs similar operations to successfully decode the ${\tf{{v}}}_{12}$- and ${\tf{{v}}}_{22}$-symbols that are intended to it.

\noindent The analysis of the sum SDoF that is allowed by the described coding scheme can be obtained by proceeding as in the proof of Theorem~\ref{theorem-sum-sdof-region-mimo-x-channel-with-asymmetric-output-feedback-and-delayed-csi}, to show that $2M^2(2M-N)$ symbols are transmitted securely to Receiver 1 over a total of $T_1+2T_2+T_3=2(4M^2-3MN+N^2)$ channel uses, thus yielding $d_{11}+d_{21}= M^2(2M-N)/{(4M^2-3MN+N^2)}$ sum SDoF at this receiver. Similar reasoning and algebra shows that $d_{12}+d_{22}= M^2(2M-N)/{(4M^2-3MN+N^2)}$ sum SDoF for Receiver 2. This concludes the proof of Theorem~\ref{theorem-sum-sdof-region-mimo-x-channel-with-asymmetric-output-feedback}.
\end{IEEEproof}

\vspace{.5em}
The analysis so far reflects the utility of both output feedback and delayed CSIT that are provided to both transmitters in terms of SDoF. However, the models that we have considered so far are \textit{symmetric} in the sense that both transmitters see the same degree of output feedback and delayed CSI from the receivers. The relative importance of output feedback and delayed CSIT depends on the studied configuration. In what follows, it will be shown that, in the symmetric model of Theorem~\ref{theorem-sum-sdof-region-mimo-x-channel-with-asymmetric-output-feedback} one can replace the asymmetric output feedback that is provided to one transmitter with delayed CSIT given to the other transmitter without diminishing the achievable sum SDoF region. 

\vspace{.5em}
\begin{figure}
\psfragscanon
\begin{center}
\psfrag{W}[][l][.8]{{\tc{blue}{$W_{11}$}},$W_{12}$}
\psfrag{X}[][l][.8]{{\tc{blue}{$W_{21}$}},$W_{22}$}
\psfrag{Y}[][r][.8]{\:\:\:\:\:\:{\tc{blue}{$\hat{W}_{11},\hat{W}_{21}$}}}
\psfrag{Z}[][r][.8]{\:\:\:\:\:\:{$\hat{W}_{12},\hat{W}_{22}$}}
\psfrag{E}[c][c][.7]{\tc{red}{$W_{12}$}}
\psfrag{F}[c][c][.7]{\tc{red}{$W_{22}$}}
\psfrag{G}[c][c][.7]{\tc{red}{$W_{11}$}}
\psfrag{I}[c][c][.7]{\tc{red}{$W_{21}$}}
\psfrag{N}[c][c]{\tc{CadetBlue}{\tf{Tx}$_1$}}
\psfrag{O}[c][c]{\tc{NavyBlue}{\tf{Tx}$_2$}}
\psfrag{P}[c][c]{\tc{CadetBlue}{\tf{Rx}$_1$}}
\psfrag{S}[c][c]{\tc{NavyBlue}{\tf{Rx}$_2$}}
\psfrag{A}[c][c]{$M$}
\psfrag{B}[c][c]{$M$}
\psfrag{C}[c][c]{$N$}
\psfrag{D}[c][c]{$N$}
\psfrag{H}[c][c]{${{\tf{H}}}$}
\psfrag{Q}[c][c]{$({{\tf{y}}}_1^{n-1},{{\tf{H}}}^{n-1})$}
\psfrag{R}[c][c]{$({{\tf{y}}}_2^{n-1},{{\tf{H}}}^{n-1})$}
\psfrag{J}[c][c]{\:\:${{\tf{x}}}_1$}
\psfrag{K}[c][c]{\:\:${{\tf{y}}}_1$}
\psfrag{L}[c][c]{\:\:${{\tf{x}}}_2$}
\psfrag{M}[c][c]{\hspace{-.2em}\:\:${{\tf{y}}}_2$}
\includegraphics[width=0.6\linewidth]{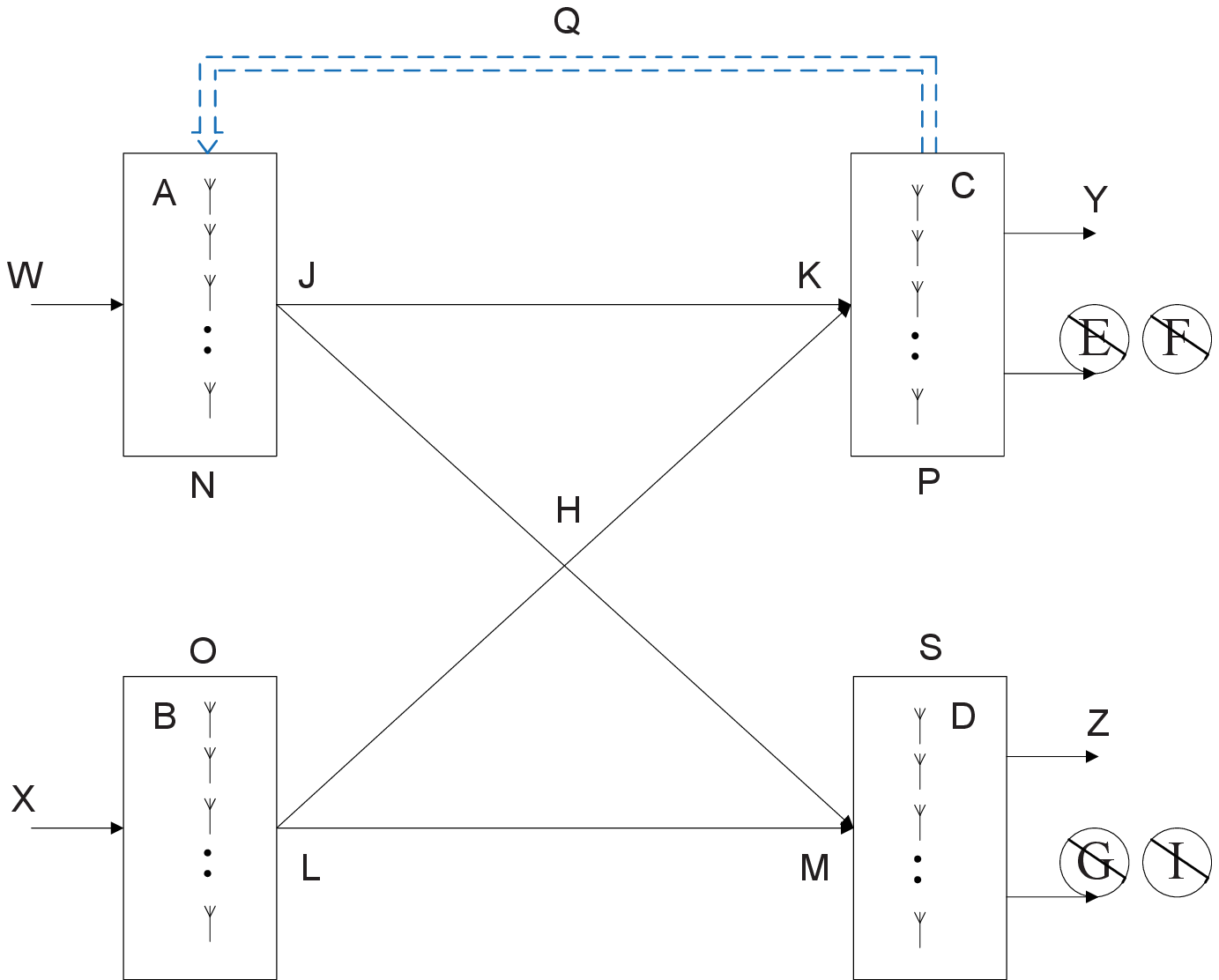}
\end{center}
\caption{MIMO X-channel with asymmetric output feedback and delayed CSIT with security constraints.}
\psfragscanoff
\label{model_asym}
\end{figure}

\begin{remark}
Investigating closely the coding scheme of Theorem~\ref{theorem-sum-sdof-region-mimo-x-channel-with-asymmetric-output-feedback}, it can be seen that the key ingredient in the achievability proof is that, at the end of the third phase, each of the side information output vector $\tilde{\bf{y}}_{2}^{(2)}$ that is required by Receiver 1 to successfully decode the symbols that are intended to it and the side information output vector $\tilde{\bf{y}}_{1}^{(3)}$ that is required by Receiver 2 to successfully decode the symbols that are intended to it be learned by \textit{exactly} one of the transmitters\footnote{By opposition, in the coding scheme of Theorem~\ref{theorem-sum-sdof-region-mimo-x-channel-with-asymmetric-output-feedback-and-delayed-csi}, both side information output vectors have been learned by both transmitters at the end of Phase 3, as we mentioned previously.}. In the coding scheme of Theorem~\ref{theorem-sum-sdof-region-mimo-x-channel-with-asymmetric-output-feedback}, the side information output vectors  $\tilde{\bf{y}}_{1}^{(3)}$ and $\tilde{\bf{y}}_{2}^{(2)}$ are learned by distinct transmitters at the end of Phase 3. The above suggests that the inner bound of Theorem~\ref{theorem-sum-sdof-region-mimo-x-channel-with-asymmetric-output-feedback} will also remain achievable if these side information output vectors are both learned by the \textit{same} transmitter. Figure~\ref{model_asym} shows a variation model that is asymmetric in the sense that asymmetric output feedback and delayed CSI are provided only to Transmitter 1. In this model, by means of the output feedback and delayed CSI from Receiver 1, Transmitter 1 can learn \textit{both} side information output vectors $(\tilde{\bf{y}}_{1}^{(3)},\tilde{\bf{y}}_{2}^{(2)})$ (See the analysis of Phase 2 in the coding scheme of Theorem~\ref{theorem-sum-sdof-region-mimo-x-channel-with-asymmetric-output-feedback-and-delayed-csi}). Taking this into account, it is easy to show that the inner bound of Theorem~\ref{theorem-sum-sdof-region-mimo-x-channel-with-asymmetric-output-feedback} is also achievable for the model shown in Figure~\ref{model_asym}. 
\end{remark}

\vspace{.5em}

\begin{proposition}
For the model with asymmetric output feedback and delayed CSI provided only to Transmitter 1 shown in Figure~\ref{model_asym}, an inner bound on the sum SDoF region is given by Theorem~\ref{theorem-sum-sdof-region-mimo-x-channel-with-asymmetric-output-feedback}.
\end{proposition}

\section{MIMO X-Channels Without Security Constraints}\label{secVI}

In this section, we consider an $(M,M,N,N)$--MIMO X-channel \textit{without} security constraints. We show that the main equivalences that we established in the previous sections continue to hold.
\vspace{.5em}
\begin{theorem}\label{theorem-sum-dof-region-mimo-x-channel-with-asymmetric-output-feedback-and-delayed-csi}
The sum DoF region $\mc C^{\text{sum}}_{\text{DoF}}$ of the two-user $(M,M,N,N)$--MIMO X-channel with asymmetric output feedback and delayed CSIT is given by the set of all non-negative pairs $(d_{11}+d_{21},d_{12}+d_{22})$ satisfying
\begin{align}
\frac{d_{11}+d_{21}}{\min(2M,2N)}+\frac{d_{12}+d_{22}}{\min(2M,N)} &\leq 1 \nonumber\\
\frac{d_{11}+d_{21}}{\min(2M,N)}+\frac{d_{12}+d_{22}}{\min(2M,2N)} &\leq 1.
\label{linear-equations-corner-points-theorem-sum-dof-region-mimo-x-channel-with-asymmetric-output-feedback-and-delayed-csi}
\end{align}
\end{theorem}
\vspace{.5em}
\begin{IEEEproof}
The converse proof follows immediately from the DoF region of a two-user MIMO BC with delayed CSIT~\cite[Theorem 2]{vaze_broadcast} in which the transmitter is equipped with $2M$ antennas and the receivers are equipped with $N$ antennas each. The proof of the direct part follows by a coding scheme that can be obtained by specializing that of Theorem~\ref{theorem-sum-sdof-region-mimo-x-channel-with-asymmetric-output-feedback-and-delayed-csi} to the setting without security constraints, and that we only outline briefly here. First, note that the region of Theorem~\ref{theorem-sum-dof-region-mimo-x-channel-with-asymmetric-output-feedback-and-delayed-csi} is fully characterized by the corner points $(\min(2M,N),0)$, $(0,\min(2M,N))$ and the point $P$ given by the intersection of the lines defining the equations in~\eqref{linear-equations-corner-points-theorem-sum-dof-region-mimo-x-channel-with-asymmetric-output-feedback-and-delayed-csi}. It is not difficult to see that the corner points $(\min(2M,N),0)$ and $(0,\min(2M,N))$ are achievable without feedback and without delayed CSIT, as the system is equivalent to coding for a MIMO multiple access channel for which the achievability follows from straightforward results. We now outline the achievability of the point $P$. If $2M \leq N$, the point $P=(M,M)$ is clearly achievable. If $N \leq 2M \leq 2N$, the achievability of the point $P=(2NM/(2M+N),2NM/(2M+N))$ can be obtained by modifying the coding scheme of Theorem~\ref{theorem-sum-sdof-region-mimo-x-channel-with-asymmetric-output-feedback-and-delayed-csi}, essentially by ignoring Phase 1. Note that, at the end of the transmission, $2MN(2M-N)$ symbols are sent to each receiver over $2T_2+T_4=(2M-N)(2M+N)$, i.e., a sum DoF of $2MN/(2M+N)$ for each. In the case in which $2M \geq 2N$, one can use the coding scheme of the previous case with each transmitter utilizing only $N$ antennas.
\end{IEEEproof}
\vspace{.5em}
\begin{remark}\label{equivalence-mimo-x-with-feedback-delayed-csi-mimo-delayed-csi}
The sum DoF region of Theorem~\ref{theorem-sum-dof-region-mimo-x-channel-with-asymmetric-output-feedback-and-delayed-csi} is same as the DoF region of a two-user MIMO BC in which the transmitter is equipped with $2M$ antennas and each receiver is equipped with $N$ antennas, and delayed CSIT is provided to the transmitter \cite[Theorem 2]{vaze_broadcast}. Thus, similar to Theorem~\ref{theorem-sum-sdof-region-mimo-x-channel-with-asymmetric-output-feedback-and-delayed-csi}, Theorem~\ref{theorem-sum-dof-region-mimo-x-channel-with-asymmetric-output-feedback-and-delayed-csi} shows that, in the context of no security constraints as well, the distributed nature of the transmitters in the MIMO X-model with a symmetric antenna configuration does not cause any loss in terms of sum DoF. This can be seen as a generalization of \cite[Theorem 1]{TMPS12a} in which it is shown that the loss is zero from a total DoF perspective. 
\end{remark}
\vspace{.5em}
\begin{remark}
Like for the setting with secrecy constraints, it can be easily shown that the sum DoF region of the $(M,M,N,N)$--MIMO X-channel with symmetric output feedback is also given by Theorem~\ref{theorem-sum-dof-region-mimo-x-channel-with-asymmetric-output-feedback-and-delayed-csi}. 
\end{remark}

\section{Numerical Examples}\label{secVII}

In this section, we illustrate the results of the previous sections (i.e., Theorems~\ref{theorem-sum-sdof-region-mimo-x-channel-with-asymmetric-output-feedback-and-delayed-csi}, ~\ref{theorem-sum-sdof-region-mimo-x-channel-with-symmetric-output-feedback}, ~\ref{theorem-sum-sdof-region-mimo-x-channel-with-asymmetric-output-feedback} and ~\ref{theorem-sum-dof-region-mimo-x-channel-with-asymmetric-output-feedback-and-delayed-csi}) through some numerical examples. We also include comparisons with some previously known results for the MIMO X-channel  without security constraints and with different degrees of CSIT and output feedback.
 
\begin{figure}
\psfragscanon
\begin{center}
\psfrag{data1}[l][l][.45]{\hspace{0em}{Sum SDoF with asymmetric output feedback and delayed CSIT, $M = 2, N = 3$}}
\psfrag{data2}[l][l][.45]{\hspace{0em}{Sum SDoF with asymmetric output feedback and delayed CSIT, $M = 4, N = 4$}}
\psfrag{data3}[l][l][.45]{\hspace{0em}{Sum SDoF with asymmetric output feedback and delayed CSIT, $M = 1, N \ge 2M$}}
\psfrag{data5}[l][l][.45]{\hspace{0em}{Sum DoF with asymmetric output feedback and delayed CSIT, $M = 2, N = 3$}}
\psfrag{data6}[l][l][.45]{\hspace{0em}{Sum DoF with asymmetric output feedback and delayed CSIT, $M = 4, N = 4$}}
\psfrag{data4}[l][l][.45]{\hspace{0em}{Sum DoF with asymmetric output feedback and delayed CSIT, $M = 1, N \ge 2M$}}
\psfrag{o}[c][c][.75]{\hspace{3em}$(2,2)$}
\psfrag{n}[c][c][.75]{\hspace{3em}$(\frac{3}{4},\frac{3}{4})$}
\psfrag{m}[c][c][.75]{\hspace{3em}$(\frac{8}{3},\frac{8}{3})$}
\psfrag{p}[c][c][.75]{\hspace{3em}$(\frac{12}{7},\frac{12}{7})$}
\psfrag{y}[c][c][.75]{$d_{12}+d_{22}$}
\psfrag{x}[c][c][.7]{$d_{11}+d_{21}$}
\includegraphics[width=0.8\linewidth]{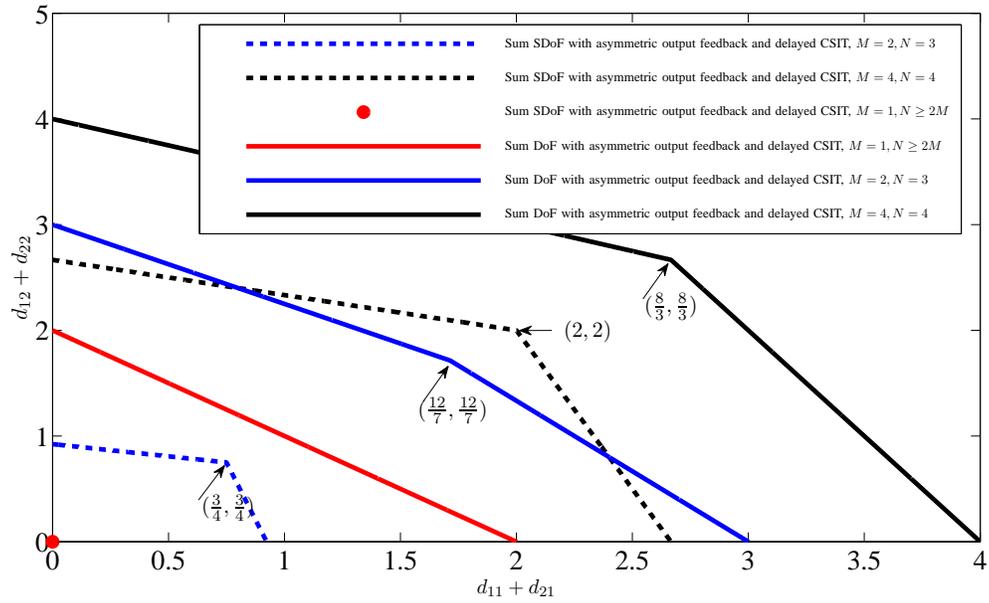}
\end{center}
\caption{Sum SDoF and sum DoF regions of the $(M,M,N,N)$--X channel with asymmetric output feedback and delayed CSIT, for different antennas configurations.}
\psfragscanoff
\label{figure-price-of-secrecy-sum-sdof}
\end{figure}

\begin{figure}
\psfragscanon
\begin{center}
\psfrag{data1}[l][l][.45]{\hspace{0em}{Sum SDoF with only asymmetric feedback, $M = 2, N = 3$}}
\psfrag{data2}[l][l][.45]{\hspace{0em}{Sum SDoF with asymmetric feedback and delayed CSIT, $M = 2, N = 3$}}
\psfrag{data3}[l][l][.45]{\hspace{0em}{Sum SDoF with asymmetric feedback and delayed CSIT, $M = N = 4$}}
\psfrag{data4}[l][l][.45]{\hspace{0em}{Sum SDoF with asymmetric feedback and delayed CSIT, $2M \le N $}}

\psfrag{data5}[l][l][.45]{\hspace{0em}{Sum SDoF with only asymmetric feedback, $M= N = 4$}}
\psfrag{data6}[l][l][.45]{\hspace{0em}{Sum SDoF with only asymmetric feedback, $2M \le N $}}
\psfrag{n}[c][c][.75]{\hspace{2em}$(\frac{3}{4},\frac{3}{4})$}
\psfrag{m}[c][c][.75]{\hspace{-2em}$(2,2)$}
\psfrag{o}[c][c][.75]{\hspace{-2em}$(\frac{4}{7},\frac{4}{7})$}
\psfrag{p}[c][c][.75]{\hspace{18em}Performance loss due to unavailability of delayed CSIT}
\psfrag{y}[c][c][.75]{$d_{12}+d_{22}$}
\psfrag{x}[c][c][.7]{$d_{11}+d_{21}$}
\includegraphics[width=0.8\linewidth]{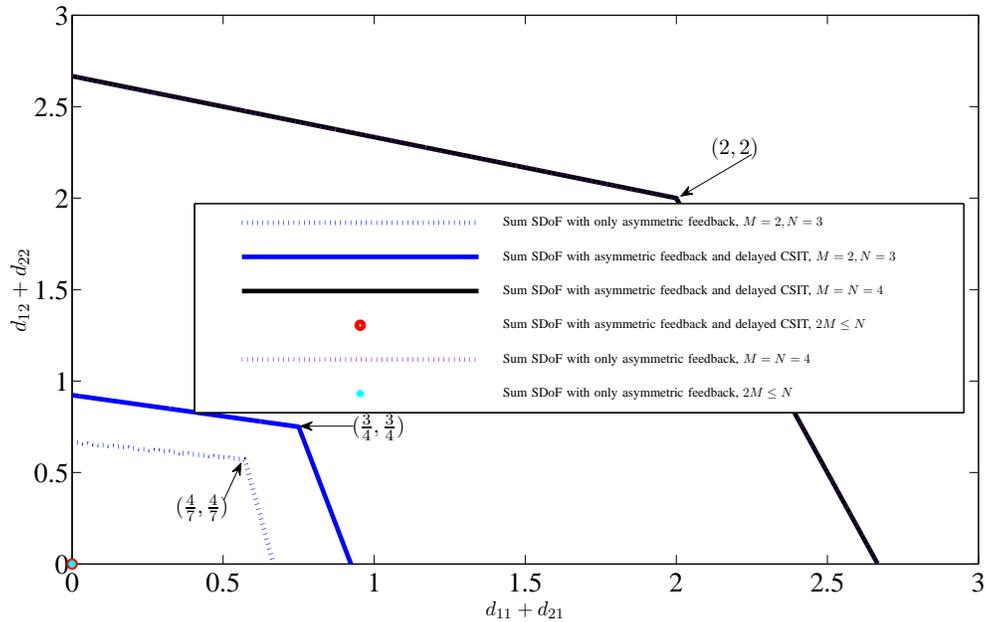}
\end{center}
\caption{Sum SDoF region of the $(M,M,N,N)$--X channel with different degrees of output feedback and delayed CSIT, for some antennas configurations.}
\psfragscanoff
\label{figure-price-of-lack-of-csi-sum-sdof}
\end{figure}

\begin{figure}
\psfragscanon
\begin{center}
\psfrag{data1}[l][l][.5]{\hspace{0em}{Total SDoF with asymmetric feedback and delayed CSIT~\eqref{total-sdof-mimo-x-channel-with-asymmetric-output-feedback-and-delayed-csi-bc-with-delayed-csi}}}
\psfrag{data2}[l][l][.45]{\hspace{0em}{Total SDoF with asymmetric feedback and no delayed CSIT [Theorem~\ref{theorem-sum-sdof-region-mimo-x-channel-with-asymmetric-output-feedback}]}}
\psfrag{data4}[l][l][.5]{\hspace{0em}{Total DoF with asymmetric feedback and delayed CSIT~\cite[Theorem~1]{TMPS12a} }}
\psfrag{data3}[l][l][.45]{\hspace{0em}{Total DoF with delayed CSIT and no feedback~\cite[Theorem~1]{akbar}}}
\psfrag{y}[c][c][.75]{$d_{11}+d_{21}+d_{12}+d_{22}$}
\psfrag{x}[c][c][.75]{Number of transmit antennas $M$ at each transmitter}
\includegraphics[height=0.5\linewidth, width=0.8\linewidth]{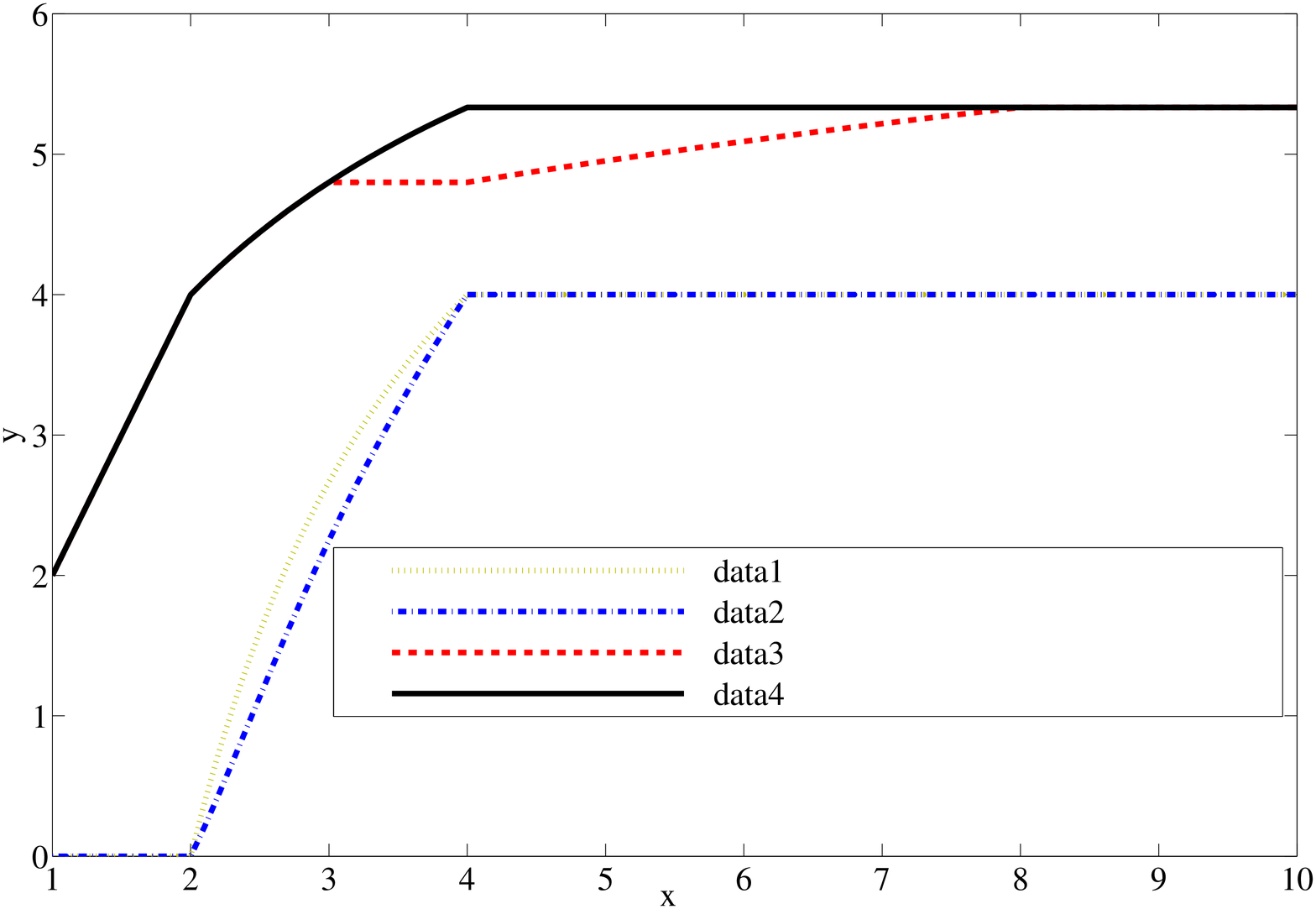}
\end{center}
\caption{Total secure degrees of freedom of the MIMO $(M,M,N,N)$--X channel, as a function of the number of transmit antennas $M$ at each transmitter, for a fixed number $N=4$ of receive antennas at each receiver.}
\psfragscanoff
\label{figure-price-of-secrecy-total-sdof}
\end{figure}

Figure~\ref{figure-price-of-secrecy-sum-sdof} illustrates the optimal sum SDoF of the  $(M,M,N,N)$--MIMO X-channel with asymmetric output feedback and delayed CSIT given by Theorem~\ref{theorem-sum-sdof-region-mimo-x-channel-with-asymmetric-output-feedback-and-delayed-csi}, for different  values of the transmit- and receive antennas. For comparison reasons, Figure~\ref{figure-price-of-secrecy-sum-sdof} also shows the optimal DoF of the same model, i.e., $(M,M,N,N)$--MIMO X-channel with asymmetric output feedback and delayed CSIT, but without security constraints, as given by Theorem~\ref{theorem-sum-dof-region-mimo-x-channel-with-asymmetric-output-feedback-and-delayed-csi}. The gap that is visible in the figure illustrates the rate loss that is caused asymptotically, in the signal-to-noise ratio, by imposing security constraints on the  $(M,M,N,N)$--MIMO X-channel with asymmetric output feedback and delayed CSIT. Thus, it can be interpreted as the \textit{price for secrecy} for the model that we study.

Figure~\ref{figure-price-of-lack-of-csi-sum-sdof} shows the inner bound of Theorem~\ref{theorem-sum-sdof-region-mimo-x-channel-with-asymmetric-output-feedback}, for different antennas configurations. As we mentioned previously, although the optimality of the inner bound of Theorem~\ref{theorem-sum-sdof-region-mimo-x-channel-with-asymmetric-output-feedback} is still to be shown, the loss in terms of secure degrees of freedom that is visible in the figure for $N \leq 2M \leq 2N$ sheds light on the role and utility of providing delayed CSI to the transmitters from a secrecy viewpoint. For $M \geq N$, however, the lack of delayed CSIT does not cause any loss in terms of secure degrees of freedom in comparison with the model with output and delayed CSIT of Theorem~\ref{theorem-sum-sdof-region-mimo-x-channel-with-asymmetric-output-feedback-and-delayed-csi}. 

Figure~\ref{figure-price-of-secrecy-total-sdof} depicts the evolution of the total secure degrees of freedom of the $(M,M,N,N)$--MIMO X-channel with asymmetric output feedback and delayed CSIT as function of the number of transmit-antennas $M$ at each transmitter, for a given number of receive-antennas at each receiver $N=4$. The figure also shows the total secure degrees of freedom with only asymmetric output feedback provided to the transmitters (obtained from Theorem~\ref{theorem-sum-sdof-region-mimo-x-channel-with-asymmetric-output-feedback}), as well as the total DoF without security constraints~\cite[Theorem 1]{TMPS12a} (which can also be obtained from Theorem~\ref{theorem-sum-dof-region-mimo-x-channel-with-asymmetric-output-feedback-and-delayed-csi}). Furthermore, the figure also shows the total DoF of the MIMO X-channel with only delayed CSIT, no feedback and no security constraints~\cite{akbar}.

\section{Conclusion}\label{secVIII}

In this paper, we study the sum  SDoF region of a two-user multi-input multi-output X-channel with $M$ antennas at each transmitter and $N$ antennas at each receiver. We assume perfect CSIR, i.e., each receiver has perfect knowledge of its channel. In addition, all the terminals are assumed to know the past CSI; and there is a noiseless asymmetric output feedback at the transmitters, i.e., Receiver $i$, $i=1,2$, feeds back its past channel output to Transmitter $i$. We characterize the optimal sum SDoF region of this model. We show that the sum SDoF region of this MIMO X-channel with asymmetric output feedback and delayed CSIT is \textit{same} as the  SDoF region of a two-user MIMO BC with $2M$ transmit antennas and $N$ antennas at each receiver and delayed CSIT. The coding scheme that we use for the proof of the direct part follows through an appropriate extension of that by Yang \textit{et al.} \cite{YKPS11} in the context of secure transmission over MIMO broadcast channels with delayed CSIT. Furthermore, investigating the role of the delayed CSIT, we also study two-user MIMO X-channel models with no CSIT. In the first model, the transmitters have no knowledge of the CSI but are provided with noiseless output feedback from both receivers, i.e., \textit{symmetric output feedback}. In the second model, each transmitters is provided by only output feedback from a different receiver, i.e., \textit{asymmetric output feedback}. For the model with symmetric output feedback, we show that the sum SDoF is same as that of the MIMO X-channel with asymmetric output feedback and delayed CSIT. For the model with only asymmetric output feedback, we establish an inner bound on the allowed sum SDoF region. Next, we specialize our results to the setting without security constraints, and show that the sum DoF region of the $(M,M,N,N)$--MIMO X-channel with asymmetric output feedback and delayed CSIT is same as the DoF region of a two-user MIMO BC with $2M$ transmit antennas and $N$ antennas at each receiver and delayed CSIT. The established results emphasize the usefulness of output feedback and delayed CSIT for transmission over a two-user MIMO X-channel with and without security constraints.

\bibliographystyle{IEEEtran}
\bibliography{macsecrecy}

\begin{IEEEbiography}[{\includegraphics[width=1.1in,height=1.3in]{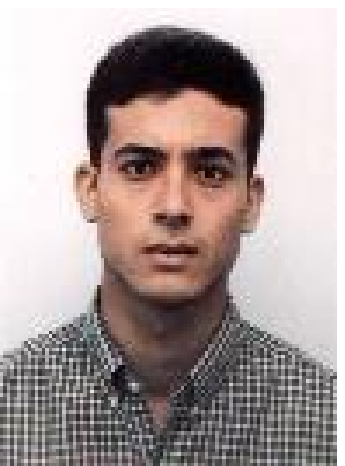}}]
{Abdellatif Zaidi}  
received the B.S. degree in Electrical Engineering from \'{E}cole Nationale Sup\'{e}rieure de Techniques Avanc\'{e}s, ENSTA ParisTech, France in 2002 and the M. Sc. and Ph.D. degrees in Electrical Engineering from \'{E}cole Nationale Sup\'{e}rieure des T\'{e}l\'{e}communications, TELECOM ParisTech, Paris, France in 2002 and 2005, respectively.

From December 2002 to December 2005, he was with the Communications and Electronics Dept., TELECOM ParisTech, Paris, France and the Signals and Systems Lab., CNRS/Sup\'{e}lec, France pursuing his PhD degree. From May 2006 to September 2010, he was at \'{E}cole Polytechnique de Louvain, Universit\'{e} catholique de Louvain, Belgium, working as a research assistant. Dr. Zaidi was "Research Visitor" at the University of Notre Dame, Indiana, USA,
during fall 2007 and Spring 2008. He is now, an associate professor at Universit\'e Paris-Est Marne-la-Vall\'ee, France.

His research interests cover a broad range of topics from signal processing for communication and multi-user information theory. Of particular interest are the problems of relaying and cooperation, network coding, interference mitigation, secure communication, coding and interference mitigation in multi-user channels, source coding and side-informed problems, with application to sensor networking and ad-hoc wireless networks. A. Zaidi is an Associate Editor of the Eurasip Journal on Wireless Communications and Networking (EURASIP JWCN).
\end{IEEEbiography}

\begin{IEEEbiography}[{\includegraphics[width=1.1in,height=1.3in]{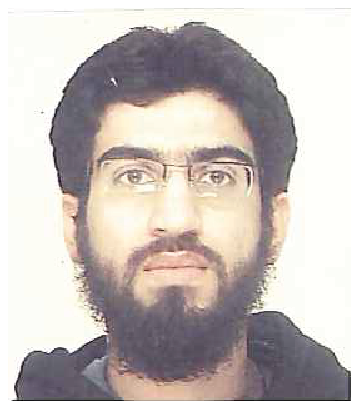}}]
{Zohaib Hassan Awan}  
 received the B.S. degree in Electronics Engineering from  Ghulam Ishaq Khan Institute (GIKI), Topi, Pakistan in 2005 and the M.S. degree in Electrical Engineering with a majors in wireless systems from Royal Institute of Technology (KTH), Stockholm, Sweden in 2008. He received the Ph.D. degree in Electrical Engineering from Universit\'{e} catholique de Louvain (UCL), Belgium in 2013. From Sept. 2013, he is with the Department
of Electrical Engineering and Information Technology, Ruhr-Universit\"{a}t Bochum, Bochum, Germany.

His current research interests include information-theoretic security, cooperative communications and communication theory.
\end{IEEEbiography}

\begin{IEEEbiographynophoto}
{Shlomo Shamai (Shitz)} received the B.Sc., M.Sc., and Ph.D. degrees in electrical engineering from the Technion---Israel Institute of Technology, in 1975, 1981 and 1986 respectively.

During 1975-1985 he was with the Communications Research Labs, in the capacity of a Senior Research Engineer. Since 1986 he is with the Department of Electrical Engineering, Technion---Israel Institute of Technology, where he is now a Technion Distinguished Professor, and holds the William Fondiller Chair of Telecommunications. His research interests encompasses a wide spectrum of topics in information theory and statistical communications.

Dr. Shamai (Shitz) is an IEEE Fellow, a member of the Israeli Academy of Sciences and Humanities and a Foreign Associate of the US National Academy of Engineering. He is the recipient of the 2011 Claude E. Shannon Award. He has been awarded the 1999 van der Pol Gold Medal of the Union Radio Scientifique Internationale (URSI), and is a corecipient of the 2000 IEEE Donald G. Fink Prize Paper Award, the 2003, and the 2004 joint IT/COM societies paper award, the 2007 IEEE Information Theory Society Paper Award, the 2009 European Commission FP7, Network of Excellence in Wireless COMmunications (NEWCOM++) Best Paper Award, and the 2010 Thomson Reuters Award for International Excellence in Scientific Research. He is also the recipient of 1985 Alon Grant for distinguished young scientists and the 2000 Technion Henry Taub Prize for Excellence in Research. He has served as Associate Editor for the Shannon Theory of the IEEE Transactions on Information Theory, and has also served twice on the Board of Governors of the Information Theory Society. He is a member of the Executive Editorial Board of the IEEE Transactions on Information Theory. 
\end{IEEEbiographynophoto}

\begin{IEEEbiography}[{\includegraphics[width=1.1in,height=1.3in]{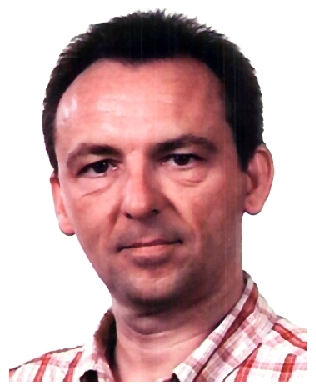}}]
{Luc Vandendorpe} (M'93-SM'99-F'06)
 was born in Mouscron, Belgium, in 1962. He received the Electrical Engineering degree (summa cum laude) and the Ph. D. degree from the Universit\'{e} catholique de Louvain (UCL) Louvain-la- Neuve, Belgium in 1985 and 1991 respectively. Since 1985, L. Vandendorpe is with the Communications and Remote Sensing Laboratory of UCL where he first worked in the field of bit rate reduction techniques for video coding. In 1992, he was a Visiting Scientist and Research Fellow at the Telecommunications and Traffic Control Systems Group of the Delft Technical University, Netherlands, where he worked on Spread Spectrum Techniques for Personal Communications Systems. From October 1992 to August 1997, L. Vandendorpe was Senior Research Associate of the Belgian NSF at UCL. Presently, he is Full Professor and Head of the Institute for Information and Communication Technologies, Electronics and Applied Mathematics of UCL.

His current interest is in digital communication systems and more precisely resource allocation for OFDM(A) based multicell systems, MIMO and distributed MIMO, sensor networks, turbo-based communications systems, physical layer security and UWB based positioning.

In 1990, he was co-recipient of the Biennal Alcatel-Bell Award from the Belgian NSF for a contribution in the field of image coding. In 2000 he was co-recipient (with J. Louveaux and F. Deryck) of the Biennal Siemens Award from the Belgian NSF for a contribution about filter bank based multicarrier transmission. In 2004 he was co-winner (with J. Czyz) of the Face Authentication Competition, FAC 2004. L. Vandendorpe is or has been TPC member for numerous IEEE conferences (VTC Fall, Globecom Communications Theory Symposium, SPAWC, ICC) and for the Turbo Symposium. He was co-technical chair (with P. Duhamel) for IEEE ICASSP 2006.

He was an editor of the IEEE Trans. on Communications for Synchronization and Equalization between 2000 and 2002, associate editor of the IEEE Trans. on Wireless Communications between 2003 and 2005, and associate editor of the IEEE Trans. on Signal Processing between 2004 and 2006. He was chair of the IEEE Benelux joint chapter on Communications and Vehicular Technology between 1999 and 2003. He was an elected member of the Signal Processing for Communications committee between 2000 and 2005, and between 2009 and 2011, and an elected member of the Sensor Array and Multichannel Signal Processing committee of the Signal Processing Society between 2006 and 2008. Currently, he is the Editor in Chief for the Eurasip Journal on Wireless Communications and Networking. L. Vandendorpe is a Fellow of the IEEE.
\end{IEEEbiography}
\end{document}